# Student's Dropout Risk Assessment in Undergraduate Course at Residential University

*Sweta Rai*
*Banasthali Univeristy*

# ABSTRACT

*Student dropout prediction is an indispensable for numerous intelligent systems to measure the education system and success rate of any university as well as throughout the university in the world. Therefore, it becomes essential to develop efficient methods for prediction of the students at risk of dropping out, enabling the adoption of proactive process to minimize the situation. Thus, this research work propose a prototype machine learning tool which can automatically recognize whether the student will continue their study or drop their study using classification technique based on decision tree and extract hidden information from large data about what factors are responsible for dropout student. Further the contribution of factors responsible for dropout risk was studied using discriminant analysis and to extract interesting correlations, frequent patterns, associations or casual structures among significant datasets, Association rule mining was applied.*

*In this study, the descriptive statistics analysis was carried out to measure the quality of data using SPSS 20.0 statistical software and application of decision tree and association rule were carried out by using WEKA data mining tool. Based on the application of association rule, the highest support value 0.68 was recorded for dropout mainly because of the personal problem. On the other hand ID3 decision tree algorithm was found best classifier with 98% accuracy whereas, discriminant function analysis correctly classified 99.1% of original grouped cases and 98.6% of cross-validated grouped cases.*

*The main reason recorded for dropout of students at this residential university were personal factor ( illness & homesickness), Educational factors (learning problems & difficult courses, change of Institution with present goal and low placement rate) and institutional factors (campus environment, too many rules in hostel life and poor entertainment facilities).*

# TABLE OF CONTENTS

# LIST OF FIGURES

## FIGURES

# LIST OF TABLES

## TABLES

# CHAPTER 1

# INTRODUCTION

We live in the information- era, accumulating data is easy and storing it inexpensive. Today the amount of stored information increases day by day in all areas. Unfortunately, as the amount of machine readable information increases, the ability to understand and make use of it does not keep pace with its growth. Stored data can be used to generate useful information for decision making. The data mining can be applied in various real life applications like market analysis, education, and scientific exploration, etc. [18].

If data is rich in quality and quantity then reliable information hidden in the data can be generated. To generate relevant information from the data which have stored in the repositories over the years, Data mining tools and techniques is used. Data mining is an automatic data analysis process that helps users and administrators to discover and extract patterns from stored data. The use of data mining technique to analyze an educational database is absolutely expected to be great benefit to the higher educational institutions.

Education is one of the social factors whereby gender disparity is reflected. The number and proportion of educated females is very low. As the grade level of education increases, the number of female students starts to decline. Consequently, higher education remains the level of learning where females are less represented both as students and staff [52]. The very few women that are fortunate enough to join higher learning institutions can be characterized by lower academic performance and higher forced withdrawal. Female education in India has got momentum after independence. There has been slow development in technical fields, but during the recent past, the continuation of female students in technical as well as in almost every field has envisaged higher status, still due to several factors there has been comparatively lesser percentage of female education as per national statistics conducted by HRD, Govt. of India. Hence, to develop a strong base of interest among female population in the country to prosecute higher education breaking the barrier of gender psychology towards which Banasthali University has played significant role. Since inception in 1935, the university has played major role in educating girl students from primary level to higher studies being undertaken from students coming from different parts of the country. As per need of the present education and employment scenario, the university has restructured its traditional educational system and introduced professional courses in the field of Engineering, Law, Management, Information & Communication Technology, Applied and pure science and Teacher's Education at undergraduate and postgraduate level. However the reasons for still shortfall in number by female students are a matter of study from various socio-economic considerations which is of great significance being undertaken both from academic and national interest under present study. Since Banasthali University is promoting mainly female education, it is more appropriate to conduct such study at this university where female students from composite strata come for high aspiration.

The purpose of this dissertation is to study the student's dropout risk assessment and causes of dropout at undergraduate level using data mining tool and techniques to assist the student dropout program on campus. Information like grade in High School, grade in Senior Secondary, student's family income, qualification of parent's etc. were collected from the student's residing in university campus, to predict the students drop out rate who need special attention.

## 1.1 Data Mining

 **"Data mining"**, often known as Knowledge Discovery in Databases (KDD), refers to mining knowledge from immense amount of data [18]. Data mining techniques are used to operate on huge amount of data to discover hidden patterns and relationships helpful for decision making. While data mining and KDD are frequently treated as synonyms, actually data mining is a part of the knowledge discovery process. The data mining defined as "the non trivial process of identifying valid, novel, previously unknown, potentially useful information, and ultimately understandable patterns from data in database" [53]. Finding a useful patterns in data are known by different names in different communities (e.g. Knowledge extraction, Information discovery, data dredging, data archeology, Information harvesting, data/pattern analysis and business intelligence) [53].

### 1.1.1 Data Mining Task

➢ **Description:** describe the dataset in a concise and summary manner and presents interesting data into human interpretable or understandable format. e.g. Clustering and Association Rule etc.

➢ **Prediction:** constructs one or a set of models, perform inference on the dataset and attempt to predict unknown or future values of other variables of interest. e.g. Classification and Regression Analysis etc.

### 1.1.2 Data Mining Functionalities

Functionalities of data mining are as follows-

➢ **Characterization**

Data characterization is a summarization of the general characteristics or features of objects in a target class of data**.** Relevant data to a user- specified class are collected by a database query and run through a summarization module to extract the essence of the data at different level of abstractions. For example, to study the characteristics of software products whose sales increased by 10% in the last year, one can collect the data related to such product by executing an SQL query. There are several methods for effective data summarization and characterization such as roll-up, drill –up, drill-down, slice, and dice.

➢ **Discrimination**

Data discrimination is a comparison of the general features of target class data objects with the general features of objects from one or a set of contrasting class. For example, compare the general feature of software products whose sale increased by 10% in the last year with those whose sales decreased by at least 30% during the same period. Discrimination descriptions expressed in rule form are referred to as discriminant rules.

➢ **Association Analysis**

Association analysis is the discovery of association rules. It studies the frequency of items take place together in transactional databases based on threshold called support and confidence. Association analysis is commonly used for market basket analysis. The discovery of association rule can help retailers to develop marketing strategies by gaining into which items are frequently purchase together.

For example- age (X,"20-29")^F_income (X,"10-20k") => dropout (X, "Yes") [Support= 2%, confidence=60%]

Rule implies that 2%(support) are 20-29 year of age with an family income of 10-20k generally drop their study and 60% probability(confidence) that a student in this age and income group will withdraw from their study.

➤ **Classification and Prediction**

Classification analyzes a set of training data and builds a model for each class based on the features in the data. This model is used to classify new objects and also known as supervised learning. Derived model may be represented in various forms, such as classification (IF- THEN) rules, decision trees, and neural network etc. A decision tree is a flow –chart like tree structure where each node denotes a test on an attribute value and each branch represents an outcome of the test. A neural network is a collection of linear threshold units that can be trained to distinguish objects of different classes. Classification can be used to predicting the class label of data objects. Classification predicts categorical labels (or discrete values), prediction models continuous- valued function. For example, a classification model may be built to categorize bank loan application as either safe or risky, while predication model may be built to predict the expenditure of potential customer equipment given their income and occupation.

➤ **Clustering**

Clustering is an unsupervised learning, in which the class labels of the training samples are not known. Cluster is a collection of data objects that are similar to on one another. Similarity can be expressed by distance functions, specified by use experts. A good clustering method produces high quality clusters to ensure that inter- cluster (object of different class) similarity is low and the intra-cluster (object in a same class) similarity is high.

➤ **Evolution and Deviation Analysis**

Evolution and deviation analysis pertain to the study, trend of object whose behavior changes over time. Evolution analysis models evolutionary trends in data, where as Deviation analysis, considers differences between measured values and expected values.

## 1.2 Educational Data Mining

Educational organizations are one of the important parts of our society and playing a vital role for growth and development of any nation. Educational data mining is the application of data mining. It is an emerging interdisciplinary research area that deals with the development of methods to explore data originating in an educational context. Educational data mining is an emerging trend, designed for automatically exploring the unique types of data from large repositories of educationally related data. Quite often, this data is extensive, fine- grained, and precise.

The Educational Data Mining community defines EDM as follows: Educational data mining is an emerging discipline, concerned with developing methods for exploring the unique types of data that come from educational settings, and using those methods to better understand students, and the settings in which they learn. Data mining is the field of discovery novel and potentially useful information from extraction huge amount of data. We know that large amount of data stored in database, so in order get the required data and to find the hidden relationship, different data mining techniques are used such as classification, prediction, association rule, outlier detection etc. EDM focuses on collection, archiving, and analysis of data related to student's learning and assessment.

EDM include analysis (evaluation/ exploration) of educational processes including admission, alumni relations, course selections, predicting drop out student, student's success rate, course success rate, performance evaluation of student, learning behavior of students, list of course taken by the student, when the student selected or changed his or her academic major, finding which tasks, courses etc are difficult/ easy for which student's, finding elective courses often taken by student's.

## 1.3 Need of Current Research

The earlier prediction of dropout student is challenging task in the higher education. Data analysis is one way to scale down the rate of dropout students and increase the enrollment rate of students in the university. It is fact that the number of student dropout quite often in the first year of graduation especially in the first semester. The rate of student's dropout in the residential university depends on the educational system adopted by the University. The needs of current research are as follows:

➢ Predicting dropout students at an early stage of the degree program help management not only to concentrate more on the bright students but also to apply more efforts in developing programs for the weaker ones in order to improve their progress while attempting to avoid student dropouts.

➢ The generated knowledge will be quite useful for understanding the problem in better way and to have a proper planning or decision to scale down the dropout rate.

➢ This study is quite useful for better planning and implementation of education program and infrastructure to increase the enrollment rate of students in particular courses provided by the university.

## 1.4 Social Implication of this Research

The larger number of dropout in the higher education has serious consequences for our society. Dropouts experiences high level of unemployment and receive lower earning than graduates [54]. Dropouts are also more likely than graduates to become dependent on welfare, engage in illegal activities, and experience health and affective problems [54, 55, 13]. Finally, high rate of dropping out of higher education create a negative momentum for youths in a society, particularly during difficult economic periods when a high school diploma does not guarantee a job. Such momentum may foster even higher dropout rates in the future. To overcome these issues this research work will be quite helpful. To reduce the rate of dropout or predict student who need special attention, we need to focus on parental support, personal and family factor, environment of campus, and quality of education.

The findings of this case study focus an outline to the policymakers with an overview of research about the dropout problem and the best strategies for building an early warning system that can signal which factors / institutions are most in need of controlling dropout problem. On the other hand, the study will also be helpful for the guardian of the students to understand the internal and the external reasons of dropout and to give them an idea about their role to minimize dropout at different levels of students. An effective measure to control dropout will significantly improve our overall education system and the nation as a whole will be benefited from it.

## 1.5 Objective

- ➢ To study the strength of relationship between attributes
- ➢ Analyze the affects of independent variables influencing graduation and dropout rates in higher education and indicates which variable are important in explaining a dropout student
- ➢ Develop a classification model using decision tree induction algorithm and classifier rules to predict whether student will graduate or not using the historic data
- ➢ Finding association of various factor leading to students dropout at higher education in residential university, where discovering of pattern or association helps in effective decision making
- ➢ To study the dropout rate and causes of students in higher education at residential university

# CHAPTER 2

# REVIEW OF LITERATURE

The several authors have been worked out in the area of educational data mining at national and international level. Some of the important studies are as follows:

**Alaa el-Halees [3],** studied how data mining is useful to improve the performance of the student in higher education. For this study association rule, classification rule using decision tree was used for analysis.

**Al-Radaideh et al. [4**], applied classification data mining techniques to improve the quality of the higher education by evaluating the main attributes of students that affect the their performance. This study was used to predict the student's final grade in a course.

**Ayesha et.al. [5],** performed study on student learning behavior. For this factors like class quizzes mid and final exam assignment are studied. This study will help the tutors to reduce the ratio of drop out and improve the performance level of students.

**Bharadwaj and Pal [6],** used the decision tree method for classification to evaluate performance of student's. The objective of their study is to discover knowledge that describes students' performance in end semester examination. This study was quite useful for identifying the dropout's student in earlier stage and students who need special attention and allow the teacher to provide appropriate advising.

**Bharadwaj and Pal [7]**, conducted study on the student performance based by selecting 300 students from 5 different degree college conducting BCA (Bachelor of Computer Application) course of Dr. R. M. L. Awadh University, Faizabad, India. By means of Bayesian classification method on 17 attributes, it was found that the factors like students grade in senior secondary exam, living location, medium of teaching, mother's qualification, students other habit, family annual income and student's family status were highly correlated with the student academic performance.

**Boero, Laureti & Naylor [9],** they found that gender (males have a higher probability of dropping out relative to the reference group of females) is one of the principal determinants of the probability of dropping out and age has a significant positive effect.

**Bray [10]**, in his study on private tutoring and its implications, observed that the percentage of students receiving private tutoring in India was relatively higher than in Malaysia, Singapore, Japan, China and Sri Lanka. It was also observed that there was an enhancement of academic performance with the intensity of private tutoring and this variation of intensity of private tutoring depends on the collective factor namely socio-economic conditions.

**Cesar et al. [11],** proposed a recommendation system based to help students to make decisions related to their academic track.

**Chandra and Nandhini [12],** used the association rule mining analysis to identifies students' failure patterns. The main objective of their study is to identify hidden relationship between the failed courses and suggests relevant causes of the failure to improve the low capacity students' performances.

**D'Mello [14]** studied on bored and frustrated student.

**El-Halees [15],** proposed a case study that used educational data mining to analyze students' learning behavior. The objective of his study is to show how useful data mining can be used in higher education to improve student's performance. They applied data mining techniques to discover relevant information from large database such as association rules and classification rules using decision tree, clustering and outlier analysis.

**Fadzilah and Abdullah [16],**applied data mining techniques to enrollment data. Descriptive and predictive approaches were used. Cluster analysis was used to group the data into clusters based on their similarities. For predictive analysis, Neural Network, Logistic regression, and the Decision Tree have been used. After evaluating these techniques, Neural Networks classifier was found to give the highest results in term of classification accuracy.

**Hijazi and Naqvi [20]**, conducted as study on the student performance by selecting a sample of 300 students (225 males, 75 females) from a group of colleges affiliated to Punjab university of Pakistan. The hypothesis that was stated as "Student's attitude towards attendance in class, hours spent in study on daily basis after college, students family income, students mother's age and mother's education are significantly related with student performance" was framed. By means of simple linear regression analysis, it was found that the factors like mother's education and student's family

**Oladipupo and oyelade [33]**, perform study using association rule data mining technique to identify student's failure patterns. They take a total number of 30 courses for 100 levels and 200 levels. Their study focuses on constructive recommendation, curriculum structure and modification in order to improve student's academic performance and trim down failure rate.

**Pandey and Pal [34]**, conducted study on the student performance based by selecting 60 students from a degree college of Dr. R. M. L. Awadh University, Faizabad, India. By means of association rule they find the interestingness of student in opting class teaching language.

**Pathom *et al.* [35],** proposed a classifier algorithm for building Course Registration Planning Model (CRPM) from historical dataset. The algorithm is selected by comparing the performance of four classifiers include Bayesian Network, C4.5, Decision Forest, and NBTree. The dataset were obtained from student enrollments including grade point average (GPA) and grades of undergraduate students. As a result, the NBTree was the best of the four classifiers. NBTree was used to generate the CRPM, which can be used to predict student class of GPA and consider student course sequences for registration planning.

**Ramasubramanian et.al.[36],** predict aspects of higher education students. In this paper they analyze that one of the biggest challenges that higher education faces today is predicting the behavior of students. Institutions would like to know, something about the performances of the students group wise. He proposed a problem to investigate the performances of the students when the large data base of Students information system (SIS) is given. Generally students' problems will be classified into different patterns based on the level of students like normal, average and below average. In this paper we attempt to analyze SIS database using rough set theory to predict the future of students.

**Ramaswami and Bhaskaran [37],** have constructed a predictive model called CHAID with 7-class response variable by using highly influencing predictive variables obtained through feature selection so as to evaluate the academic achievement of students at higher secondary schools in India. Data were collected from different schools of Tamilnadu, 772 students' records were used for CHAID prediction model construction. As a result, set of rules were extracted from the CHAID prediction model and the efficiency was found. The accuracy of the present model was compared with other models and it has been found to be satisfactory.

**Romero and Ventura [38],** survey the relevant studies carried out in the field of education. They have described the types of users, types of educational environments and the data they provide. Also they have explained in their work the common tasks in the educational environment that have been resolved through data mining techniques.

**Romero & Ventura [39],** provided a survey of educational data mining from 1995-2005 and Baker & Yacef (2009) extended their survey covering the latest development until 2009. There are an increasing number of data mining applications in education, from enrollment management, graduation, academic performance, gifted education, web-based education, retention and other areas (Nandeshwar & Chandhari, 2009). In this section we will review only research where the main focus is on study outcome, i.e. successful or unsuccessful course completion. Based on his open learning model Kamber (1995) stated that entry, i.e. background characteristics are not good predictors of final outcomes because they are just a starting point and there are other factors that may contribute to the difficulties student will have to deal with during his/her study.

**Shaeela Ayesha, Tasleem Mustafa, Ahsan Raza Sattar, and M. Inayat Khan [40]** applied K-mean clustering to analyze learning behavior of students which will help the tutor to improve the performance of students and reduce the dropout ratio to a significant level.

**Shannaq et al. [41],** applied the classification as data mining technique to predict the numbers of enrolled students by evaluating academic data from enrolled students to study the main attributes that may affect the students' loyalty (number of enrolled students). The extracted classification rules are based on the decision tree as a classification method, the extracted classification rules are studied and evaluated using different evaluation methods. It allows the University management to prepare necessary resources for the new enrolled students and indicates at an early stage which type of students will potentially be enrolled and what areas to concentrate upon in higher education systems for support.

**Sun [43],** the student learning result evaluation system is an essential tool and approach for monitoring and controlling the learning quality. This paper conducted a research on student learning result based on data mining. With this model in practice, student learning can become more energetic, more interesting, more challenging, and more suited to the times and this research paper will help to understand student learning evaluation system to generate theories.

**Superby [45]** conducted a study to investigate to determine the factors to be taken into account we will use a model adapted from that of Philippe Parmentier (1994). In other words the idea is to determine if it is possible to predict a decision variable using the explanatory variables which we retained in the model.

**Tissera *et al*. [46],** presented a real-world experiment conducted in an ICT educational institute in Sri Lanka, by analyzing students' performance. They applied a series of data mining task to find relationships between subjects in the undergraduate syllabi. They used association rules to identify possible related two subjects' combination in the syllabi, and apply correlation coefficient to determine the strength of the relationships of subject combinations identified by association rules. As a result, the knowledge discovered can be used for improving the quality of the educational programs to plain English to be used as a factor to be considered by the managerial system to either support their current decision makings or help them to set new strategies and plan to improve their decision making procedures. The main idea of this analysis is organized into the DM-HEDU guideline proposed by the authors, which targets the superior advantages of data mining in higher learning institution. The authors have presented several projects of using data mining in higher education.

**Vandamme, Meskens & Superby [47],** used decision trees, neural networks and linear discriminant analysis for the early identification of three categories of students: low, medium and high risk students. Some of the background information (demographics and academic history) of the first-year students in Belgian French-speaking universities were significantly related to academic success. Those were: previous education, number of hours of mathematics, financial independent Early Prediction of Student Success: Mining Students Enrolment Data 650 attendance ence, and age, while gender, parent's education and occupation, and marital status were not significantly related to the academic success. However, all three methods used to predict academic success did not perform well. Overall the correct classification rate was 40.63% using decision trees, 51.88% using neural networks and the best result was obtained with discriminant analysis with overall classification accuracy of 57.35%.

**Walters and Soyibo [48],** conducted a study to determine Jamaican high school students' (population n=305) level of performance on five integrated science process skills with performance linked to gender, grade level, school location, school type, student type, and socio-economic background (SEB). The results revealed that there was a positive significant relationship between academic performance of the student and the nature of the school.

**Warapon in [49],** who presented the use of data mining techniques, particularly classification, to supports high school students in selecting undergraduate programs. Warapon proposed a classification model to give guidelines to students, especially, for the undergraduate programs for making possible better academic plans. The decision tree technique was applied to determine which major is best suitable for students.

**Woodman [50]** found for courses in the mathematics and computing faculty at the Open University in UK, by using the binary logistic regression, that the most significant factors to whether students passed, failed or dropped out, were marks for the first assignment, the number of math courses passed in the previous two years, the course level, the points the course is worth and the occupation group of the student. This was the most parsimonious model, but in the model which includes all 25 potential predictors other variables such as ethnicity (ranked as 7th according to its relative importance), education (8th), age group (9th), course level (11th), disability (18th) and gender (22nd) were also significant. However, one of the problems with the logistic regression is that in large samples any difference, may lead to conclusion that the factor is significant when in fact that is not the case. Using the same methodological approach with data available at new student registration in the UK Open University Simpson (2006) found that the most important factor is the course level, followed by credit rating of a course, previous education, course programme, socio-economic status, gender and age.

**Yadav, Bharadwaj and Pal [51]**, To predict the students performance they obtained the university students data like attendance, class test, seminar and assignment marks using three algorithms ID3, C4.5 and CART.

# CHAPTER 3

# METHODOLOGY

Success percentage rate of any institute can be improved by knowing the reasons for dropout student. In present study, information on various parameters was collected through a structured questionnaire on personal interview basis from a composite sample of 220 students of undergraduate courses (BCA and B.Tech) in computer science / information technology of University. Predicting the students dropout status whether they continue to their study or not, needs lots of parameters such as personal, academic record, social, environmental, etc. variables are necessitated for the effective prediction.

Since the present study is in relation to classify the various quantitative and qualitative factors to study the causes of dropout which belongs to the process of knowledge discovery and data mining. This information will be helpful for the management to reduce the dropout rate in campus. In order to achieve the above mentioned objectives the following steps were followed (Fig. 1):

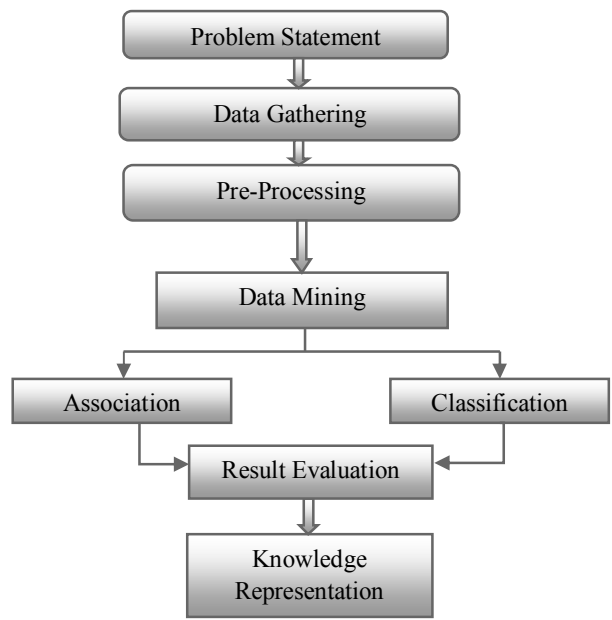

**Figure 1: Work Methodology**

## 3.1 Data Preparation

The data used in this study was prepared from the University through structured questionnaire on personal interview. The questionnaire has been constructed based on theoretical and empirical grounds about factor affecting student's performance and causes of dropout. The questionnaire included socio-demographic indicators ( Age, Date of birth, Geographical location, Marital status, Parents education, Parents occupation and Annual income), Educational factors (Performance in High school, Senior Secondary School , Location of Schooling, Type of Examination Board, Medium of Study etc.), Parental Attitudes, Causes of dropout, and Institutional factors, etc. Data was collected from 1st year undergraduate (BCA and B.Tech) students. The data format is presented in Table 1.

Before the initial visit to review the records, a coding system was created for each variable to be documented (e.g., rural=0, urban=1). The socio-demographic data, school factors, family element and attitude towards institution were investigated as a part of this research (see Appendix A for coding sheet). It was not important to document dropout status but also all withdrawal reasons for the students.

**Table 1: Student Related Variables**

| VARIABLES | DESCRIPTION | POSSIBLE VALUES |
|---|---|---|
| AGE | Age | {<18, 18-20, >20} |
| RES | Residence | {Rural, Urban} |
| N_STATE | Native state | Categorical |
| FTYPE | Family type | {Nuclear, Joint} |
| ANN | Annual Income | {Low, Medium, High, VHigh} |
| FEDU | Father's education | {Illiterate,  Sec, HSec, UG,PG} |
| MEDU | Mother's education | {Illiterate, Sec, HSec, UG, PG} |
| FOCC | Father's occupation | {Govt. service, Pvt. Service, Business, Agriculture, Retried, NA} |
| MOCC | Mother's occupation | {Govt. service, Pvt. Service, Business, HWife, NA} |
| SSCG | 10th Grade | {A=90-100%, B=80-89%, C=70-79%, D=60-69%, E=less than 60%} |
| HSCG | 12th Grade | {A=90-100%, B=80-89%, C=70-79%, D=60-69%, E=less than 60%} |
| S_LOC | location of school | {village, town city} |
| MED | Medium of schooling | {Hindi, English} |
| HSC_STREAM | Stream in higher secondary | {Math, Bio, Com, Arts, Arts(Math)} |
| C_ADMITTED | Enrolled in course | {B.Tech, BCA} |
| A_TYPE | Admission type | {Entrance, Merit, Management} |

| SAT_LEVEL | Student's satisfaction with course | {V.Satisfied, Satisfied, Not V.Satisfied, Not Satisfied} |
|---|---|---|
| C_SYLL | Syllabus of course | {V.satisfactory, satisfactory, Balanced, difficult, V.difficult, Lengthy} |
| Uni_EXPENSES | Expenses in university | {own_income, Loan, Both} |
| STRESS | Any type of stress in family | {No, Financial, illness, Other} |
| ULIK | Like University | {Yes, No} |
| UES | University Education sys. | {Excellent, V.Good, Good, Poor, V.Poor} |
| UINF | University infrastructure | {Excellent, V.Good, Good, Poor, V.Poor} |
| CURR | Extracurricular in university | {Excellent, V.Good, Good, Poor, V.Poor} |
| ENTER | Availability of entertainment in campus | {Excellent, V.Good, Good, Poor, V.Poor} |
| Self study | Time spare for study | {<1 hr, 2-3 hrs,  4-5 hrs, >6 hrs} |
| PAR_CURR | Participation in extracurricular activity | {Yes, No} |
| PLAC | Placement status | {below avg, avg, good, V.good, Excellent} |
| DROP | Withdraw from course | {Yes, Not Sure, No} |

## 3.2 Data Selection and Transformation

After collection of data, the dataset was prepared to apply the data mining techniques. Before application of prescribed model, data preprocessing was applied to measure the quality and suitability of data. In this step only those attributes were selected which were needed for data mining. For this, remove missing values; smoothing noisy data, selection of relevant attribute from database or removing irrelevant attributes, identifying or remove outlier values from data set, and resolving inconsistencies of data. Some of the irrelevant parameters was removed from database such as ID, age, date of birth, category, marital status, state of domicile, mother tongue, religion, the gender field containing only one value- female, because university concerns only female students, the marital status field containing one value- unmarried, etc. A categorical variable is constructed based on the numeric parameter percentage in secondary and higher secondary school. A grade scale is used for evaluation of student performance at school. "Garde A" students are considered those who have a percentage greater than 85, "Grade B"- in the range between 75 and 85, "Grade C"- in the range between 65 and 75, and "Grade D" in the range below 65. A four level scale is used in the family annual income. "VHigh" annual income are considered those who have income greater than 6 lakhs, "High" annual income range between 4 lakhs and 6 lakhs, "Medium" annual income range between 2 lakhs and 4 lakhs and "Low" annual income in the range below 2 lakhs. A categorical target variable "Dropout status" is constructed based on the view of respondents; it has two possible values- "Yes" (students who are completely decided to withdraw from their course) and "No" (students who are want to continue their study).

The final dataset used for the study contains 220 instances (183 in the "No" category and 37 are in "Yes" category) each described with 34 attributes (1 output and 33 input variables), nominal and numeric. The study is limited to the student data for undergraduate. Finally, the pre-processed data were transformed into a suitable format to apply data mining techniques.

## 3.3 Data Analysis Techniques

In this study, one quantitative and the other qualitative data analysis techniques have to be employed using statistical methods and data mining methods.

### Statistical Methods

The data collected were analyzed using SPSS statistical software to measure the quality of data based on descriptive statistics, crosstab and statistical test. Examine each variable by using cross tab. The entire student was examined to understand the reasons of withdrawal.

### Data Mining Methods

After proper collection, scrutiny and transformation of data using appropriate measures, Data mining classification and decision tree approach were applied to predict student dropout rates and dropout causes in early stage of their study either before or after completion of their first year of their study program. Classification model will be implemented by using WEKA tool. There are several classifiers available in WEKA but ID3 was used for the purpose of the study. Attribute important analysis was carried out to rank the attributes by significance using information gain. Correlation based Feature selection (CFS) using Best First searching technique and Discriminant analysis were used to rank and select the attributes that are most useful. Further association rule mining technique was used to discover relationship between seemingly unrelated variables in large database with great potential used in various real life applications including business and industry.

**Correlation-Based Feature Selection**

Feature selection is used to select a subset of input data most useful for analysis and future prediction by eliminating features, which are irrelevant of no predictive information. Features selection is use for increasing the predictive accuracy and reducing complexity of learner results [28]. In present study Correlation-Based Feature Selection (CFS) was used to find the feature subsets that are highly correlated with the class but minimal correlation between features combined with search strategy best-first search (BFS). Best First Search method starts with empty set of features and generates all possible single feature expansions. The subset with highest evaluation is chosen and expanded in the same manner by adding single features. If expanding a subset results in no improvement, the search back to the next best unexpanded subset and continues from there.

Equation for CFS"

$$M_s = \frac{K \overline{r_{cf}}}{\sqrt{K + K(K-1)\overline{r_{ff}}}}$$

Where $M_s$ is the heuristic "merit" of feature subset S containing K features, $\overline{r_{cf}}$ is the mean feature-class correlation and $\overline{r_{ff}}$ is the average feature-feature inter-correlation.

The search begins with the empty set of features, which has zero merit. The subset with highest merit found during search the search is used to reduce the dimensionality of training and test data. The search will terminate if five consecutive fully expanded subsets show no improvement over the current best subset. Reduced datasets may be passed to machine learning (ML) for building a classifier to predict the dropout student as shown in Fig. 2.

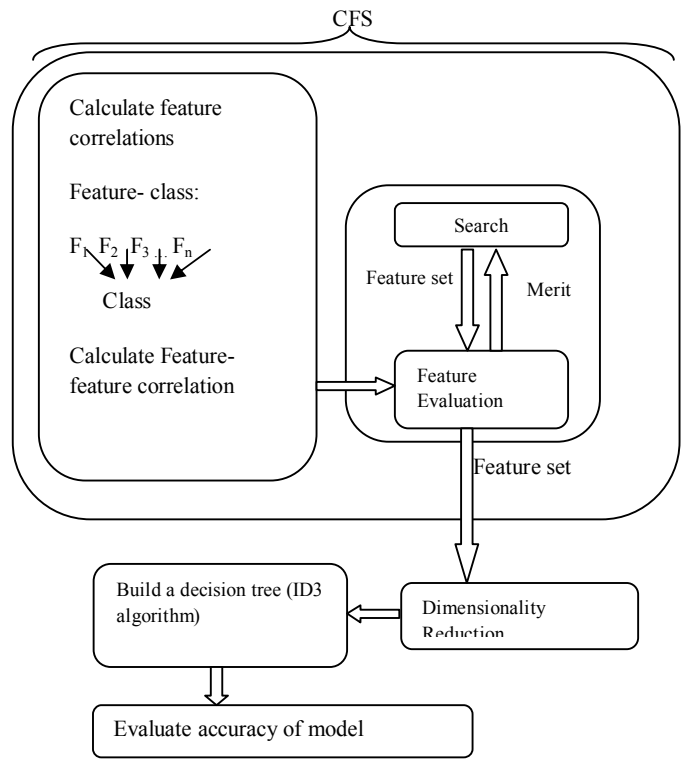

**Figure 2: Flow Chart of Feature Selection**

To build a decision tree ID3 employ a greedy approach (information theoretic measure) and correlation feature selection (CFS).

**ID3 (Iterative Dichotomizer 3)**

ID3 (Iterative Dichotomizer 3) algorithm is invented by J. Ross Quinlan in 1979. It is used for building the decision tree using information theory invented by in 1948. It builds the decision tree from top down, with no backtracking. Information Gain is used to select the best attribute for classification.

**Entropy**

It is a measure of uncertainty about a source of message. It ranges from 0 to 1. When entropy is 1 means dataset is homogenous. Entropy is calculated by formula:

$$Entropy(s) = \sum_{i=1}^{c} -P_i Log_2 P_i$$

Where, $P_i$ is the probability of S belonging to class i.

**Information Gain**

It measures the expected reduction in entropy. ID3 calculates the Gain of all attributes, and select the one with highest gain. To calculate Gain use formula:

$$Gain(S, A) = Entropy(S) - \sum_{v \in Values(A)} \frac{|S_v|}{|S|} Entropy(S_v)$$

Where, values(A) is the set of all possible values for attribute A, and $S_v$ is the subset of S for which the attribute A has value v. The attribute with highest information gain among the attributes are located as root node in the decision tree.

**Implementation of ID3 Algorithm**

**Step 1:** compute classification entropy.

**Step 2:** for each attribute, calculate information gain using classification attribute.

**Step 3:** select attribute with highest information gain.

**Step 4:** remove node attribute, for future calculation.

**Step 5:** repeat steps 2-4 until all attribute have been used.

The experiments are conducted using Weka tool. Run a CFS with search space best-first search on dataset and record the selected features. Then apply ID3 only on the selected features set and record overall accuracy by 10-fold cross-validation.

**Performance Evaluation**

To evaluate the performance of classification, this paper has adopted precision, recall and F-measure as a performance measure.

$$True\ positive = \frac{TP}{P}$$

$$False\ Negative = \frac{FP}{N}$$

$$Recall = \frac{TP}{TP + FN}$$

$$Precision = \frac{TP}{TP + FP}$$

$$F - measure = \frac{2 * Recall * Precision}{Recall + Precision}$$

**Association Rule**

For mining a frequent item set, association rule in data mining plays an important role. Association rule is defined as an implication form X => Y, where intersection of X and Y is null. For measuring the usefulness of association rules, statistical significance called support and goodness called confidence framework is used.

The Support S is used to define as the percentage of transactions in the data set, which contain the item set.

$$support(X \rightarrow Y) = \frac{tuples\ containing\ both\ X\ and\ Y}{total\ tuples}$$

The confidence is defined as a conditional probability. Confidence (X=>Y) is the ratio of number of transaction that contains XUY.

$$confidence(X \rightarrow Y) = \frac{Support\ (XUY)}{Support(X)}$$

The confidence of a rule measures the strength of the rule means correlation between antecedent and the consequent while the support measures the frequency of the antecedent and consequent together.

Apriori algorithm is used for association rule mining, was developed in 1994 by R. Agrawal and R. Srikant [2] for mining frequent itemsets for Boolean association rules. Apriori employs an iterative approach, where K itemsets are used to explore (K+1) itemsets. To generate the excellent result support and confidence measures are considered [2]. Apriori algorithm is two- step process: In first step find all item sets having support factor is greater than or equal to specified minimum support. In second step generate all association rules having the confidence factor greater than or equal to minimum confidence threshold value.

# CHAPTER 4

# RESULTS AND DISCUSSION

The methodology described in the previous chapter provides the baseline for data gathering; however the present chapter focuses on analysis and interpretation of data using statistical tools and data mining techniques. The data collected was compiled in Microsoft Excel-2007 software. The SPSS software was used for frequency distribution and descriptive statistics analysis for each response variable. After measuring the quality of data, data mining classification techniques were applied to study the factors affecting student's dropout in higher education at University level and subsequently causes for dropout was also analyzed.

## 4.1 Description of the sample

As stated from the first chapter, the aim of the study was to analyze the factors influencing dropout student and its reasons. For this study, data was collected by using questionnaires which were delivered to the respondents by hand (N=300) of BCA and B.Tech (CS/IT) courses along with approval letter of university authority and instructions of research problem. After collection of datasets, about 80 datasets were deleted due to incomplete information. The remaining (N=220) datasets were analyzed which includes information like socio- demographic factors (age, category, marital status, residential status, family status, religion), Parental status (annual income, parent's education, parent's occupation), educational profile of respondents (grade in secondary school, grade in higher secondary, location of school, medium of education, stream in higher secondary, course admitted, admission criteria, source of education expenses, study duration) and attitude of respondent to the university (course satisfaction, syllabus of course, education system, infrastructure, extracurricular activities, and teacher student relationship etc.). The findings of the research problems are as follows.

## 4.2 Descriptive statistics

**One Way Classification**

The One-way frequency distribution is the most straightforward form of analysis and one that often supplies much of the basic information need, is to tabulate results, question by question, as 'one-way tables'. Of course this does not identify which respondents produced particular combinations of responses, but this is often a first step to obtain summaries of individual variables. This is most useful to determine the typical values of the variables, checking the assumption of statistical procedures and checking the quality of data. In this section frequency distribution of 220 sample cases collected at university level was worked out to study the nature and quality of data for each single variable related with demographic, educational and institutional factors of respondents. Figure 3 demonstrates the level of study of survey participants:

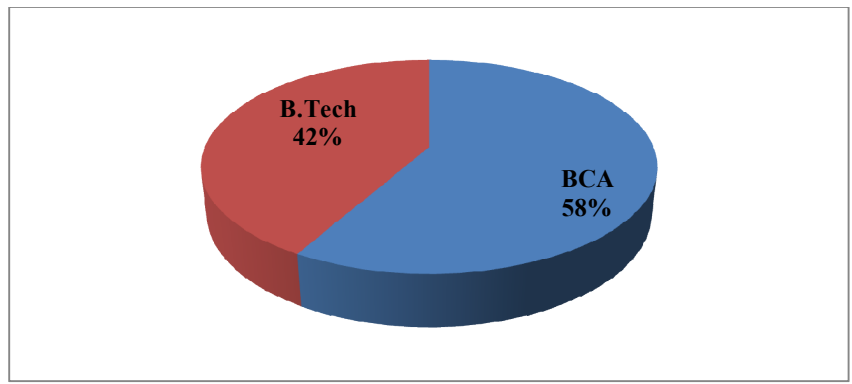

**Figure 3: Respondent's enrolled for the Study Programme**

More than half of the respondents (58%) were enrolled for Bachelor of Computer Application and 42% of respondents were enrolled for Bachelor of Technology.

**Demographic Profile of Respondents**

The demographic issues explored in this study were Age, Category, Residential status and Family status. Table 2 shows the demographic profile of the participants. The age range for the participants was from 16 to 23 years. More than half of the respondents (131) belong to age group 18 -20 years; whereas about 70 respondents are below 18 years and remaining 19 respondents are 20 years & above. In terms of category, majority of respondents (78.2%) were from General and a few of them from OBC (20.5%) and SC (1.4%) categories. The Rural – Urban distribution clearly indicate that majority of enrolled respondents (76.8%) were from urban area whereas only 23.2% respondents participation was recorded from Rural area. The low participation from rural area may be due to lack of awareness about the university, financial constraints and low level of parental attitude towards educating girls in higher technical education. About more than 90% respondents belong to Hindu religion and their mother tongue is Hindi. In context of family type distribution, the higher occurrence of students (52.3%) was recorded from nuclear family and remaining 47.7% from Joint family.

**Table 2: Demographic Profile of the Participants**

| Demographic factors | Particulars | Frequency (no. of students) | Percent |
|---|---|---|---|
| Age | Below 18 | 70 | 31.8 |
| | 18-20 | 131 | 59.5 |
| | 20 & Above | 19 | 8.6 |
| Category | General | 172 | 78.2 |
| | OBC | 45 | 20.5 |
| | SC | 3 | 1.4 |
| Marital Status | Unmarried | 220 | 100.0 |
| Residential Status | Urban | 169 | 76.8 |
| | Rural | 51 | 23.2 |
| Mother Tongue | Hindi | 209 | 95 |
| | Others | 11 | 5 |
| Religion | Hindu | 204 | 92.7 |
| | Jainism | 7 | 3.18 |
| | Sikh | 7 | 3.18 |
| | Muslim | 2 | .9 |
| Family Type | Nuclear | 115 | 52.3 |
| | Joint | 105 | 47.7 |

**Socio-economic Profile of Respondent's Parents**

It was found that both father and mother of the respondents were highly educated. It was found that 85.9% fathers were graduate and above, whereas about 71.8% mothers were graduate and above. The remaining 14.1% fathers and 28.1% mother were having qualification up to Higher Secondary level (Fig. 4). Occupational trend shown in Fig. 5 indicates that, majority of respondents' fathers (39.1%) were engaged in Govt. service and Business (42.7%) and rest of them associated with private service (14.1%) and agriculture (4.1%). Whereas, maximum number of mothers (79.5%) were house wife and few of them belong to Govt. service (10%), Pvt. Service (8.2%) and Business (2.3%). In case of annual income (Fig. 6), most of the respondents' parents (45.5%) were earning in the range of Rs. 200001 to 400000 followed by Rs.

200000 & below (29.1%) and Rs. 400001 to 600000 (16.8%) respectively. However, a few of the respondents' parents' annual income (8.6%) was recorded Rs. 600001 & above.

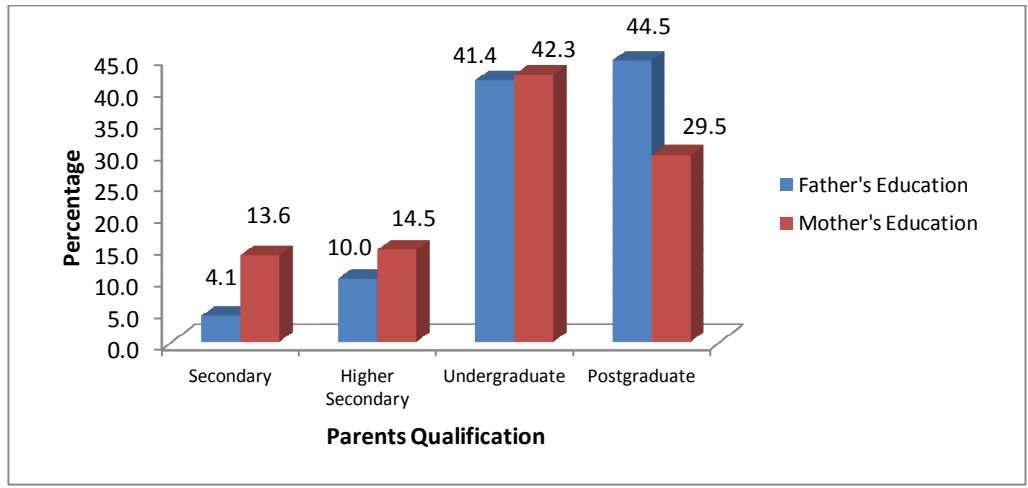

**Figure 4: Profile of Sample by Parent's Qualification**

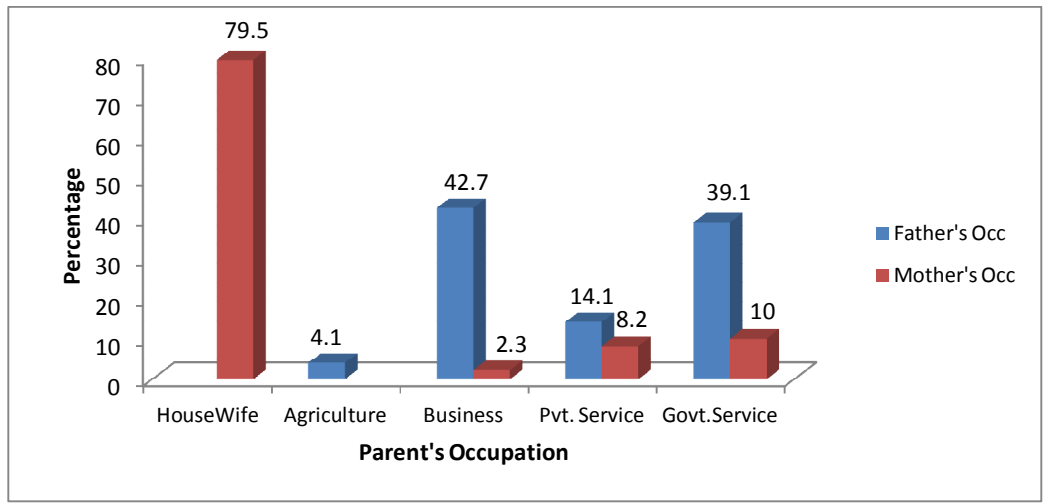

**Figure 5: Profile of Sample by Parent's Occupation**

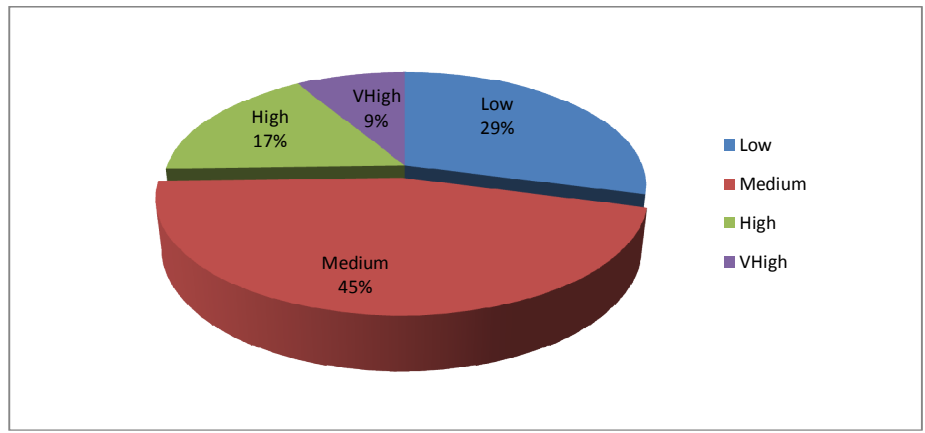

**Figure 6: Annual Income Group Respondent's Parent**

**Educational profile of the Respondents**

Table 3 shows the educational profile of the respondents related with secondary and higher secondary grade points, School location, Medium & Stream of education in HSC, course admitted after completion of HSC and graduation, admission criteria, source of education expenses and study duration respectively. About 40.9% respondents obtained grade A in SSC followed by 31.4% in grade B, 21.4% in grade C and rest 6.4% in grade D. On the other hand in HSC, 19.5% respondents secured marks in A grade, 50.5% B grade, 26.8% C grade and 3.2% D grade (Fig. 7).

Majority of the students (67.7%) admitted in HSC belongs to City area followed by Town area (24.1%) and from villages only 8.2%. Subsequently it was found that most of the respondents were studied from English medium (68.6%) and remaining from Hindi medium (31.4%). The stream in HSC indicates that maximum students have opted science with mathematics (81.4%) and few of them (18.6%) opted Commerce, Arts and Biology groups. After passing their HSC and graduation exam, the respondents applied for BCA, B.Tech at the university and admission criteria for respective courses were merit based on qualifying marks and entrance test. Based on marks basis, 41.8% were admitted in BCA course and after qualifying entrance test and 58.2% got admission in B.Tech.

In terms of educational expenses, majority of respondent (93.6%) were bearing the educational expenditure from their own source of income and about 6.4% respondent meet their expenses with the help of bank loan.

Study duration in present course of computer science & Application reveals that more than 60% students spend 2 hrs time and 38% between 4 hrs. Though the university is completely residential and educational system has been managed in such way so that girl student can manage more time for self study for better performance in Information and communication Technology courses at undergraduate and post graduate level.

**Table 3: Educational Profile of Respondents**

| Educational factors | Particulars | Frequency | Percent |
|---|---|---|---|
| SSC Grade | A (85-100) | 90 | 40.9 |
| | B (75-85) | 69 | 31.4 |
| | C (65-75) | 47 | 21.4 |
| | D (55-65) | 14 | 6.4 |
| | E (45-55) | 0 | 0.0 |
| HSC Grade | A (85-100) | 43 | 19.5 |
| | B (75-80) | 111 | 50.5 |
| | C (65-75) | 59 | 26.8 |
| | D (55-65) | 7 | 3.2 |
| | E (45-55) | 0 | 0.0 |
| School Location in HSC | Village | 18 | 8.2 |
| | Town | 53 | 24.1 |
| | City | 149 | 67.7 |
| Medium in HSC | Hindi | 69 | 31.4 |
| | English | 151 | 68.6 |
| Stream in HSC | Science(Math) | 179 | 81.4 |
| | Science(Bio) | 5 | 2.3 |
| | Commerce | 20 | 9.1 |
| | Arts(Math) | 8 | 3.6 |
| | Arts | 8 | 3.6 |
| Course admitted | BCA | 128 | 58.2 |
| | B.Tech | 92 | 41.8 |
| Admission criteria | Marks basis | 128 | 58.2 |
| | Entrance test | 92 | 41.8 |
| Source of education expenses | Own Income | 206 | 93.6 |
| | Bank Loan | 14 | 6.4 |
| Study duration | 2 hrs | 136 | 61.8 |
| | 4 hrs | 84 | 38.2 |

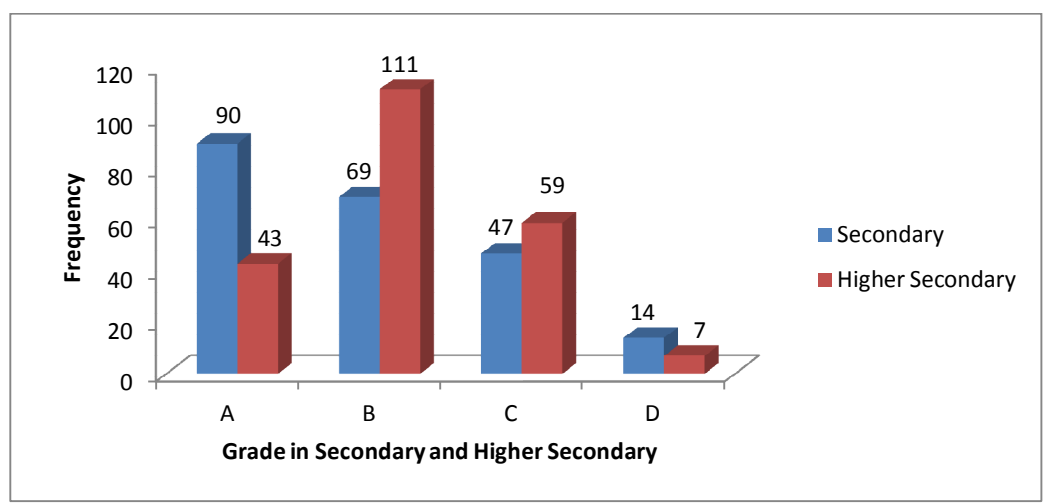

**Figure 7: Respondent's Educational Performance at School Level**

**Reaction of Respondents against Statements**

The response to the question are presented in Table 4, shows about reaction of students against statements about course satisfaction, type of syllabus, education system, infrastructure, entertainment, extracurricular activities, and placement status of the university. The maximum learners (25.9%) very satisfied and (55.5%) satisfied with course opted at the university. In terms of course syllabus, 35.5% learners reported satisfactory, 30.9% lengthy, and 20% balanced. About 75% students ranked the educational system, Infrastructure and extracurricular activities of the university as good and very good. In terms of placement rate, about 40% respondents reported good and very good. On other hand 57.3% were in favor of average placement rate. The majority of learners (90.9%) were in favor to have better relation between teacher and student at university campus. About 69.1% respondents were cope with pressure at university by one or other reasons which may be due to academic, personal or environmental factors. During course of study programme, about 58.6% students were free from any stress condition, whereas 21.8% students were facing financial stress and few of them suffer due to illness etc (Fig. 8).

**Table 4: Reaction of Students against Statements**

| Institutional Factors | Response | Frequency | Percent |
|---|---|---|---|
| Course satisfaction | Not Very Satisfied | 41 | 18.6 |
| | Satisfied | 122 | 55.5 |
| | Very Satisfied | 57 | 25.9 |
| Syllabus of the course | Difficult | 30 | 13.6 |
| | Lengthy | 68 | 30.9 |
| | Satisfactory | 78 | 35.5 |
| | Balanced | 44 | 20.0 |
| Education system | Excellent | 40 | 18.2 |
| | Very Good | 79 | 35.9 |
| | Good | 93 | 42.3 |
| | Poor | 8 | 3.6 |
| Infrastructure | Excellent | 30 | 13.6 |
| | Very Good | 59 | 26.8 |
| | Good | 113 | 51.4 |
| | Poor | 18 | 8.2 |
| Extracurricular activities | Excellent | 81 | 36.8 |
| | Very Good | 102 | 46.4 |
| | Good | 37 | 16.8 |
| Entertainment status | Excellent | 25 | 11.4 |
| | Very Good | 43 | 19.5 |
| | Good | 82 | 37.3 |
| | Poor | 70 | 31.8 |
| Participation in Extracurricular activities | No | 76 | 34.5 |
| | Yes | 144 | 65.5 |
| Cope with Pressure at the university | No | 68 | 30.9 |
| | Yes | 152 | 69.1 |
| Teacher-Student relation | No | 20 | 9.1 |
| | Yes | 200 | 90.9 |
| Placement status | Average | 126 | 57.3 |
| | Good | 47 | 21.4 |
| | Very Good | 39 | 17.7 |
| | Excellent | 8 | 3.6 |

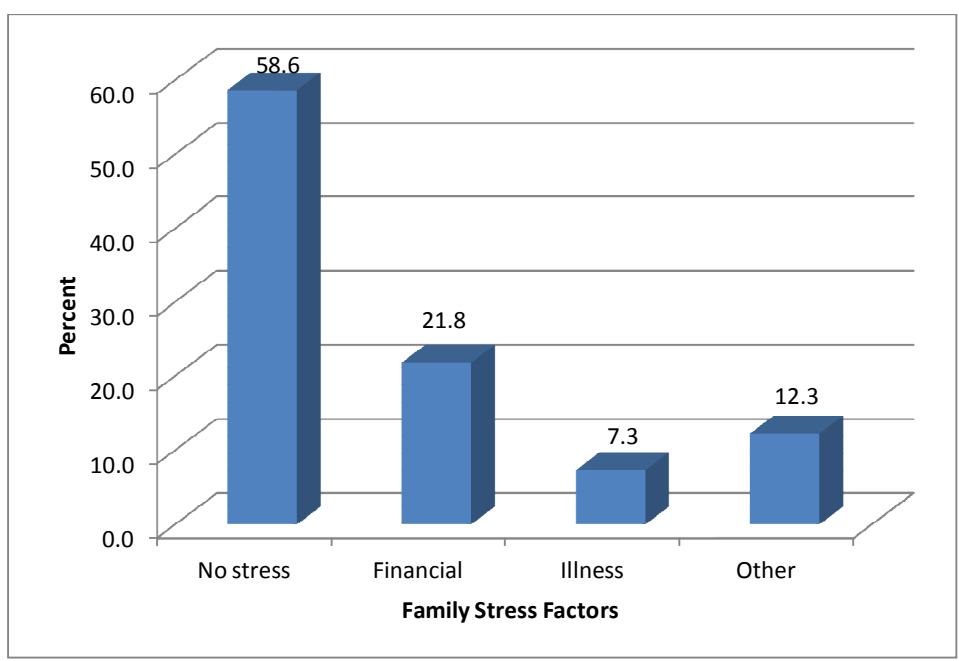

**Figure 8: Family Experiencing Stress Factors**

Table 5 shows the response of the respondents related to source of information collected about university programme and their decision to join this university. Majority of students came to know about this university through their parents and friend's which indicates that parents were well versed about the reputation of the university in terms of education, security system and financial expenditure. About more than 75% respondents joined this university by their own choice and after enrollment they also like the university in terms of academic and environment.

**Table 5: Source of Information and Preference about University**

| Factors | Particulars | Frequency | Percent |
|---|---|---|---|
| How you came to know about this University? | Internet<br>Parents<br>Newspaper<br>Friends<br>Others | 30<br>144<br>53<br>155<br>51 | 6.9<br>33.3<br>12.2<br>35.8<br>11.8 |
| You join the university by your own choice | No<br>Yes | 52<br>168 | 23.6<br>76.4 |
| Do you like this university? | No<br>Yes | 30<br>190 | 13.6<br>86.4 |

Sample size (n): 220

## Cross Tabulation Analysis

The course wise respondents response presented in Table 6 shows non-significant relationship among the attributes that majority of respondents 56% from BCA and 54% from B.Tech groups were simply satisfied with their courses. Whereas, about 26% BCA and 25% B.Tech respondents were very satisfied with ICT courses at University level. Only 17% BCA and 21% B.Tech respondents were not satisfied with the opted courses it may be due to professional interest, institutional and personal problems etc.

**Table 6: Course Wise Respondents Response**

| ICT Course | Response | | | Total |
|---|---|---|---|---|
| | Not Satisfied | Satisfied | Very Satisfied | |
| BCA | 22<br>(17.2) | 72<br>(56.2) | 34<br>(26.6) | 128<br>(100.0) |
| B.Tech | 19<br>(20.7) | 50<br>(54.3) | 23<br>(25.0) | 92<br>(100.0) |
| Total | 41<br>(18.6) | 122<br>(55.5) | 57<br>(25.6) | 220<br>(100.0) |

Chi-square value is $0.430^{NS}$ and parenthesis values are in percent

The response obtained by respondents in relation to structure of syllabus framed by university for BCA and B.Tech courses in Table 7 were significant at 1% level (chi-square = $11.91^{**}$). The satisfactory response was recorded for BCA was 33.6% and for B.Tech (38%). On ther other hand response against balance syllabus was recorded 17.2% and 23.9% by BCA and B.Tech respondents. However a majority of respondents 28.9% (BCA) and 33.7% (B.Tech) expressed their view that the course syllabus of ICT was lengthy.

**Table 7: Students' Opinion about Syllabus of ICT Courses**

| ICT courses | Syllabus of Course | | | | Total |
|---|---|---|---|---|---|
| | Difficult | Lengthy | Satisfactory | Balanced | |
| BCA | 26 (20.3) | 37 (28.9) | 43 (33.6) | 22 (17.2) | 128 (100.0) |
| B.Tech | 4 (4.3) | 31 (33.7) | 35 (38.0) | 22 (23.9) | 92 (100.0) |
| Total | 30 (13.6) | 68 (30.9) | 78 (35.5) | 44 (20.0) | 220 (100.0) |

Chi-square value is 11.91 significant at 1% level, and parenthesis values are in percent

The association between university education system vs. ICT courses in Table 8 was found significant at 1% level (chi-square = $20.73^{**}$). The respondent of BCA (38.3%) and B.Tech (32.6%) group categories very good about university education System. Whereas about 42.3% respondents including BCA and B.Tech graded about education system of the university in relation to ICT programmes as good. Remaining 25.8% BCA and 7.6% B.Tech respondents graded as excellent education system. Overall majority of respondents like the educational system of the residential university because they use to get all required facilities at one place under better environment.

**Table 8: Course Wise Students' Reaction about University Education System**

| | University Education System | | | | Total |
|---|---|---|---|---|---|
| | Poor | Good | Very Good | Excellent | |
| BCA | 1 (0.8) | 45 (35.2) | 49 (38.3) | 33 (25.8) | 128 (100.0) |
| B.Tech | 7 (7.6) | 48 (52.2) | 30 (32.6) | 7 (7.6) | 92 (100.0) |
| Total | 8 (3.6) | 93 (42.3) | 79 (35.9) | 40 (18.2) | 220 (100.0) |

Chi-square value is 20.731, Significant at 1% level and parenthesis values are in percent

Respondents grading about university infrastructure was recorded in terms of ICT undergraduate courses in Table 9 shows significant (Chi-square = $8.40^{*}$) at 5% level. Majority of BCA (45.3%) and B.Tech (59.8%) students graded about university infrastructure as good followed by Very good grading (26.8%) and Excellent (13.6%).

**Table 9: Grading of Students about University Infrastructure for Course Opted**

| | University Infrastructure | | | | Total |
|---|---|---|---|---|---|
| | Poor | Good | Very Good | Excellent | |
| BCA | 8 (6.2) | 58 (45.3) | 41 (32.0) | 21 (16.4) | 128 (100.0) |
| B.Tech | 10 (10.9) | 55 (59.8) | 18 (19.6) | 9 (9.8) | 92 (100.0) |
| Total | 18 (8.2) | 113 (51.4) | 59 (26.8) | 30 (13.6) | 220 (100.0) |

Chi-square value is 8.402, Significant at 5% level and and parenthesis values are in percent

The response about dropout rate of the respondents was recorded in Table 10 shows significant difference (Chi-square = 11.98$^{**}$) at 1% level. Majority of respondent from BCA (75.8%) and B.Tech (93.5%) were not willing to dropout the course, whereas dropout rate just after enrollment or during first year of their study in BCA and B.Tech was found minimal i.e. in the range of 7 to 24% only. This may be due to their personal and professional problems.

**Table 10: Dropout Rates vs. Course Opted in ICT**

| | Dropout status | | Total |
|---|---|---|---|
| | No | Yes | |
| BCA | 97 (75.8) | 31 (24.2) | 128 (100.0) |
| B.Tech | 86 (93.5) | 6 (6.5) | 92 (100.0) |
| Total | 183 (83.2) | 37 (16.8) | 220 (100.0) |

Chi-square value is 11.983, significant at 1% level and parenthesis values are in percent

Dropout rate in terms of family problems was recorded in Table 11, shows significant difference (Chi-square = 13.00$^{**}$) at 1% level. Dropout rate was mainly influenced by family problems (51.4%) and remaining 48.6% may be due to other factors such as health problems and institutional environment etc.

**Table 11: Dropout Rate in Terms of Family Problem**

| Dropout | Family Problem | | Total |
|---|---|---|---|
| | No | Yes | |
| No | 142 (77.6) | 41 (22.4) | 183 (100.0) |
| Yes | 18 (48.6) | 19 (51.4) | 37 (100.0) |
| Total | 160 (72.7) | 60 (27.3) | 220 (100.0) |

Chi-square value is 13.002, significant at 1% level and parenthesis values are in percent

Dropout rates in terms of campus environment recorded in Table 12 shows significant difference (Chi-square = $39.50^{**}$) at 1% level. Now it is clearly noted that 48.6% (18) out of 37 respondents of BCA & B.Tech have dropout risk due to campus environment.

**Table 12:   Dropout Rates Vs. Campus Environment**

| Dropout | Like Campus Environment | | Total |
|---|---|---|---|
| | No | Yes | |
| No | 168 (91.8) | 15 (8.2) | 183 (100.0) |
| Yes | 19 (51.4) | 18 (48.6) | 37 (100.0) |
| Total | 187 (85.0) | 33 (15.0) | 220 (100.0) |

Chi-square value is 39.500, significant at 1% level and parenthesis values are in percent

## 4.3   Application of Discriminant Analysis

The forward stepwise method of discriminant analysis was used for classification of variables in which variables are gradually added to the model until satisfactory criteria have been met. At each step variables with highest F ratio value added to the model. The process ends when F enters value for no variable is higher than the specified F enter value [1, 2]. In this analysis following F values were used:   F-enter value is 3.84 an F-remove value is 2.71 [8].

**Table 13: Variables Entered/Removed[a,b,c,d]**

| | | Wilks' Lambda | | | | | | | |
|---|---|---|---|---|---|---|---|---|---|
| | | | | | | Exact F | | | |
| Step | Entered | Statistic | df1 | df2 | df3 | Statistic | df1 | df2 | Sig. |
| 1 | Family Stress | .492 | 1 | 1 | 218.000 | 225.503 | 1 | 218.000 | .000 |
| 2 | Participation Activity | .405 | 2 | 1 | 218.000 | 159.688 | 2 | 217.000 | .000 |
| 3 | Campus environment | .356 | 3 | 1 | 218.000 | 130.295 | 3 | 216.000 | .000 |
| 4 | SatisfactionLevel | .317 | 4 | 1 | 218.000 | 116.072 | 4 | 215.000 | .000 |
| 5 | Change goal | .287 | 5 | 1 | 218.000 | 106.127 | 5 | 214.000 | .000 |
| 6 | Family Type | .265 | 6 | 1 | 218.000 | 98.692 | 6 | 213.000 | .000 |
| 7 | residence | .247 | 7 | 1 | 218.000 | 92.251 | 7 | 212.000 | .000 |
| 8 | Family Problem | .236 | 8 | 1 | 218.000 | 85.556 | 8 | 211.000 | .000 |
| 9 | Home sickness | .225 | 9 | 1 | 218.000 | 80.225 | 9 | 210.000 | .000 |
| 10 | Self study | .220 | 10 | 1 | 218.000 | 73.935 | 10 | 209.000 | .000 |
| 11 | medium | .216 | 11 | 1 | 218.000 | 68.742 | 11 | 208.000 | .000 |
| 12 | TS Relation | .211 | 12 | 1 | 218.000 | 64.423 | 12 | 207.000 | .000 |

At each step, the variable that minimizes the overall Wilks' Lambda is entered.

a. Maximum number of steps is 66.

b. Minimum partial F to enter is 3.84.

c. Maximum partial F to remove is 2.71.

To determine whether or not there is statistically significant relationship between the independent variables and the dependent variables see Table 14. Number of function is equal to 1 less than numner of levels defined in group variable. In this analysis dependent variable "Dropout Status" has two level 0 (No) and 1 (Yes) therefore only one function is considered. Conclusions about the significance of model were made on the basis of Wilks' lambda, commonly used in discriminant analysis [40]. The discriminant function was significant as Wilks' lambda was .211 (p-value < 0.0001). Its value indicates the existence of differences between groups, that is, the fact that two groups have different means. The model is constructed by with 12 variables that have significant effect upon discrimination between groups [17]. Wilks' lambda is the portion of total variances in the discriminant score NOT explained by differences among the groups. Table 2 shows 21% of the variance is not explained by group differences. Range of Wilks' Lambda is 0 to 1. The larger the value, the smaller significance and the smaller the value, the larger the significance is ensured.

The canonical correlation coefficient .888 measures the association between the discriminant score and the set of independent variables. Like Wilks' lambda, it is an indicator of the strength of relationship between entities in the solution.

Wilks' lambda is used to test null hypothesis that the means of all the independent variables are equal across groups of the dependent variable. If the chi-square statistic corresponding to Wilks' lambda is statistically significant we conclude that there is a relationship between the dependent groups and the independent variables. Here chi-square is 329.642 and significant at .0001.

The eigenvalue assess the relative importance because they reflect the percentage of variance explained in the dependent variable and estimated value is 3.735. Larger the eigenvalue indicate that the discriminant functions is more useful in the distinguishing between groups.

**Table 14: Significance Test of Discriminant Function**

| Function | 1 |
|---|---|
| Eigenvalue | 3.735 |
| Canonical Correlation | .888 |
| Wilks' Lambda | .211 |
| Chi- square | 329.642 |
| Df | 12 |
| p-value | .000 |

Standardized discriminant coefficient is used to compare the relative importance of independent variable [17]. Table 15 shows that Family stress is highly correlated to dropout student.

**Table 15: Standardized and Structured coefficient**

| Variables | Standard coefficient | Structure coefficient |
|---|---|---|
| Residence | .336 | .100 |
| familyType | .311 | .193 |
| Medium | .198 | .102 |
| satisfactionLevel | -.327 | -.236 |
| familyStress | .774 | .526 |
| selfStudy | -.174 | -.054 |
| participationActivity | -.510 | -.312 |
| TSRelation | -.168 | -.036 |
| familyProblem | .238 | .130 |
| Homesickness | .265 | .203 |
| campusEnvironment | .329 | .242 |
| changeGoal | .238 | .266 |

Structure matrix shows the correlation for each variable with each discriminant function. Based on the structured coefficient value, it can be concluded that academic performance of students in university is mainly determine by Family stress. The student who will not graduate is determined by variables: Residence, Family type, Medium, Satisfaction level, Family stress, Self study, Participation activity, Teacher student relationship, Family Problem, Home sickness, Campus environment, Change goal. In other words, students will dropout, whose residence is urban, family type is joint, education medium is Hindi, who is not satisfied, campus environment is not so good, and who has some stress in family and the student who are not participated in any extra-activity.

**Table 16: Canonical Discriminant Function Coefficient**
**(Unstandardized coefficient)**

| Variables | Function 1 |
|---|---|
| Residence | .808 |
| familyType | .661 |
| Medium | .432 |
| satisfactionLevel | -.539 |
| familyStress | 1.058 |
| selfStudy | -.179 |
| participationActivity | -1.247 |
| TSRelation | -.582 |
| familyProblem | .547 |
| Homesickness | .985 |
| campusEnvironment | 1.013 |
| changeGoal | 1.055 |
| Constant | -.323 |

Canonical discriminant function coefficients are used to design a discriminant score function. Discriminant score is the product of the unstandardized coefficients and predicators.

$$Discriminant\ function(D)$$
$$= -.323 + (.808 * residence) + (.661 * familyType)$$
$$+ (.423 * medium) + (-.539 * satisfactionLevel)$$
$$+ (1.058 * familyStress) + (-.179 * selfStudy)$$
$$+ (-1.247 * participationActivity) + (-.582 * TSRelation)$$
$$+ (.547 * familyProblem) + (.985 * HomeSickness)$$
$$+ (1.013 * CmpsEnvironment) + (1.055 * changeGoal)$$

**Table 17: Classification Results**

| Dropout Status | | Predicted Group Membership | | Total |
|---|---|---|---|---|
| | | No | Yes | |
| Original (actual class) | No | 182 (99.5%) | 1 (0.5%) | 183 (100.0%) |
| | Yes | 1 (2.7%) | 36 (97.3%) | 37 (100.0%) |
| Cross validated (actual class) | No | 182 (99.5%) | 1 (0.5%) | 183 (100.0%) |
| | Yes | 2 (5.4%) | 35 (94.6%) | 37 (100.0%) |

99.1% of original grouped cases correctly classified
98.6% of cross-validated grouped cases correctly classified

The classification results contain two parts Original and Cross Validated. The Original classification may be biased because all cases in the analysis are classified by functions created using all cases in the sample. And another is cross validated, uses split halves method, the original dataset is split into two randomly and one half is used to develop the discriminant equation and other half is used to validate it. The classification results indicated in Table 17 reveal that about 99.1% of original grouped cases and 98.6% of cross-validated grouped cases were correctly classified into dropout status (yes and no) groups by the model. The student who will continue their study were classified with slightly better accuracy (99.5%) than the student who will completely decided to drop out at university (94.6%).

Histogram (Fig. 9) is alternative way of illustrating the distribution of the discriminant function score for each group. Graph shows that there is very minimal overlap between each group. This suggests that the function does discriminate well.

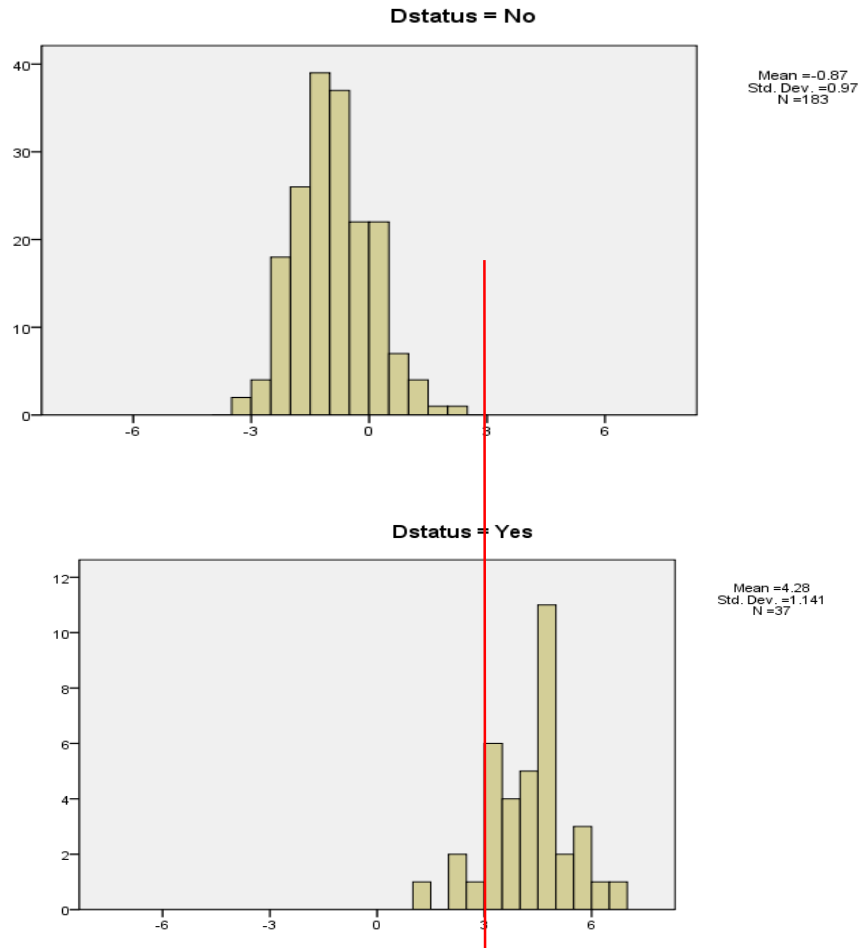

**Figure 9: Discriminant score for student who completely decided to dropout or not**

Overall, out of 33 variables under study only 12 variables such as Residence, Family type, Medium, Satisfaction level, Family stress, Self study, Participation activity, Teacher student relationship, Family Problem, Home sickness, Campus environment and Change of goal were significantly contributed to discrimination between student who will dropout or not.

## 4.4 Application of Decision Tree Induction Algorithm

The aim of this study is to select a highly correlated feature which are associated with dropout student and design a classification model for future prediction of whether student will drop the course or continue and study the cause of dropout students.

At this stage, in order to generate knowledge the 12 most representative attribute were selected from database based on correlation feature selection method (Fig. 10).

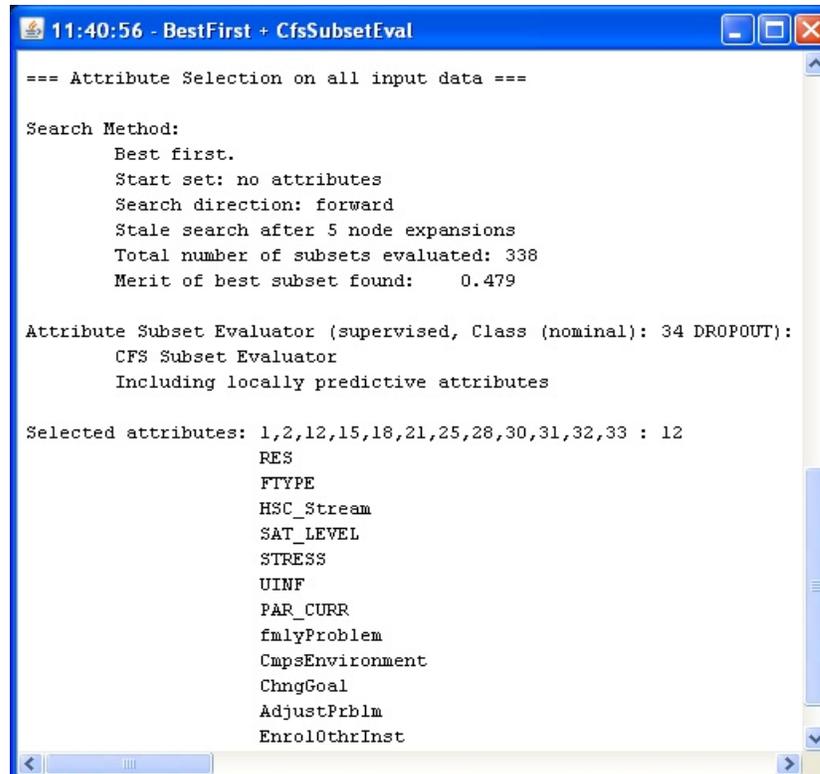

**Figure 10: Attribute Selected Using Correlation Feature Selection Background of ID3**

CFS with search strategy best-first implemented on the input data, it could be seen that subsets of 12 attributes was selected out of 34 attributes (Table 18). Then the ID3 decision tree algorithm is employed on selected subset of features and record using 10 cross- validation. Attribute with highest information gain is used as a root node (Table 19). This process goes on until all data is classified perfectly. Finally, build a decision tree using information gain value (Fig. 11). To study the actual and predicted class a confusion matrix table was used (Table 20). The dropout data set is classified into two groups Yes and No based on this 2x2 confusion matrix for id3 was constructed shows accuracy percent 98.1818 for ID3. It indicates that it is the best classifier for predicting the student who will dropout or not at the University.

**Table 18: Feature Selection based on Correlation Feature Selection**

| Variables | Description | Selected by Correlation based feature selection |
|---|---|---|
| Res | Residence | Yes |
| FType | Family Type | Yes |
| FAIn | Family annual income | |
| FEdu | Father's Education | |
| MEdu | Mother's Education | |
| FOcc | Father's Occupation | |
| MOcc | Mother's Occupation | |
| SLoc | School location of student | |
| HSG | Students grade / percentage in High School ($10^{th}$) | |
| SSG | Students grade / percentage in Senior Secondary ($12^{th}$) | |
| Med | Medium of education | |
| HSC_Stream | Stream in Senior Secondary($12^{th}$) | Yes |
| CAdm | Course Admitted | |
| AType | Admission Type | |
| SAT_Level | Satisfaction with course | Yes |
| CSyllabus | Syllabus of Course | |
| UExpenses | Parents meet the University Expenses | |
| Stress | Family experiences stress | Yes |
| LikeUni | Like this university | |
| EduU | Educational system of the university | |
| UINF | Infrastructure of university | Yes |
| ActU | Extra-Curriculum activities in university | |
| EntertU | Entertainment in university | |
| SelfStudy | Time for self study | |
| Par_Curr | Participate in extra curriculum activity | Yes |
| PlacementStatus | Placement status | |
| TSRelation | Teacher student relationship | |
| fmlyProblem | Family problem | Yes |
| Hsickness | Home sickness | |
| cmpsEnv | Campus Environment | Yes |
| chngGoal | Change of goal | Yes |
| adjustPrblm | Adjustment Problem | Yes |
| EnrolOthrInst | Enrolled in other institute | Yes |

**Table 19: Ranked Attribute with Respect to Information Gain**

| Information Gain | Attribute |
|---|---|
| 0.355 | STRESS |
| 0.1925 | PAR_CURR |
| 0.1471 | HSC_Stream |
| 0.1442 | SAT_LEVEL |
| 0.1129 | EnrolOthrInst |
| 0.1058 | ChngGoal |
| 0.1015 | CmpsEnvironment |
| 0.0953 | FTYPE |
| 0.0834 | UINF |
| 0.0513 | AdjustPrblm |
| 0.0389 | fmlyProblm |
| 0.0329 | RES |

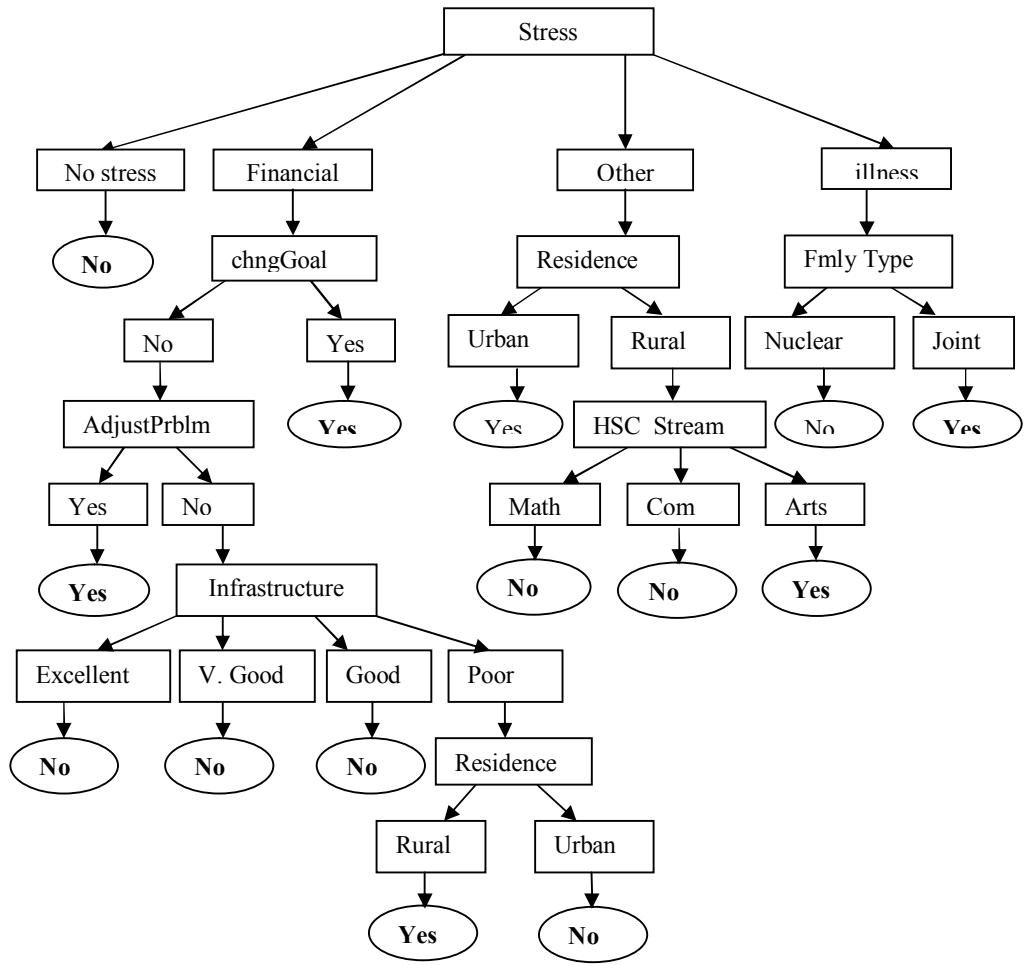

**Figure 11: Decision Tree Using ID3 Algorithm (Weka tool)**

After tree construction and confusion matrix, evaluation parameters such as Recall, F-measure, Precision and Accuracy are calculated shown in Table 22.

**Table 20: Confusion Matrix of ID3**

|  |  | Predicted class | | |
|---|---|---|---|---|
|  |  | No | Yes | Total |
| Actual class | No | 182 (TP) | 1 (FN) | 183 |
|  | Yes | 3 (FP) | 34 (TN) | 37 |
|  | Total |  |  | 220 |

$$Correctly\ Classified\ instances\ (accuracy) = \frac{TP + TN}{TP + TN + FP + FN}$$

$$= \frac{182 + 34}{220} * 100 = 98.1818\%$$

$$Incorrectly\ Classified\ instances = \frac{FP + FN}{N} = \frac{1 + 3}{220} * 100 = 1.8182\%$$

$$Misclassification\ rate\ (mean\ absolute\ error) = \frac{FN + FP}{N} = \frac{3 + 1}{220} = 0.0182$$

$$Sensitivity\ (Recall) = \frac{TP}{TP + FN} = \frac{182}{182 + 1} = 0.995$$

$$Specificity = \frac{TN}{FP + TN} = \frac{34}{3 + 34} = 0.919$$

$$Precision = \frac{TP}{TP + FP} = \frac{182}{182 + 3} = 0.984$$

$$F - measure = \frac{2 * Recall * Precision}{Recall + Precision} = \frac{2 * 0.995 * 0.984}{0.995 + 0.984} = 0.989$$

**Table 21: Results for the ID3 Decision Tree Algorithm Using 10- fold Cross Validation (Accuracy by Class)**

|  | Class | |
|---|---|---|
| **Accuracy** | No | Yes |
| TP rate | 0.995 | 0.919 |
| FP rate | 0.081 | 0.005 |
| Precision | 0.984 | 0.971 |
| Recall | 0.995 | 0.919 |
| F-measure | 0.989 | 0.944 |
| ROC area | 0.957 | 0.957 |

**Table 22 Classifier Rules**

| Rule |
|---|
| IF Stress=No THEN Dropout= No |
| IF Stress= Financial AND chngGoal=No AND AdjustPrblm= No AND infrastructure= good THEN Dropout= No |
| IF Stress= Financial AND chngGoal=No AND AdjustPrblm= No AND infrastructure= V.good THEN Dropout= No |
| IF Stress= Financial AND chngGoal=No AND AdjustPrblm= No AND infrastructure= Poor AND Residence= Urban THEN Dropout= No |
| IF Stress= Financial AND chngGoal=No AND AdjustPrblm= No AND infrastructure= Poor AND Residence= Rural THEN Dropout= No |
| IF Stress= Financial AND chngGoal=No AND AdjustPrblm= No AND infrastructure= Excellent THEN Dropout= No |
| IF Stress= Financial AND chngGoal=No AND AdjustPrblm= Yes THEN Dropout=Yes |
| IF Stress= Financial AND chngGoal=Yes THEN Dropout=Yes |
| IF Stress=Other AND Residence=Urban THEN Dropout=Yes |
| IF Stress=Other AND Residence=Rural AND HSC_Stream=Math THEN Dropout=No |
| IF Stress=Other AND Residence=Rural AND HSC_Stream=Commerce THEN Dropout=No |
| IF Stress=Other AND Residence=Rural AND HSC_Stream=Arts THEN Dropout=No |
| IF Stress=illness AND Ftype=nuclear THEN Dropout=No |
| IF Stress=illness AND Ftype=Joint THEN Dropout=Yes |

Use the classifier rules to improve student admission plan, tracking and help the students who have a high probability of dropping out including educational quality management planning of the university.

## 4.5 Application of Association Rule

In this study data was accumulated from the students who discontinue their studies. These data are analyzed using Association Rule Mining to find out the causes or factors behind their decision to discontinue their education at the university. In order to achieve the objective following steps were performed in sequential order:

### Data Set

Under data set, the reason provided by the students for dropping out of the ICT courses at university level were divided into four factors such as family problem, health related, personal problem and institutional problem listed in Table 23. The size of dataset was 37.

**Table 23: Dataset with Description of Attribute and Possible Values**

| Variable | Description | Possible value |
|----------|-------------|----------------|
| Factor 1 | Family problem | Family problem |
| Factor 2 | Health problem | illness |
| Factor 3 | Personal problem | Home sickness, marriage, peer problem, adjustment problem, change of the goal, course were difficult, course fee is high, learning problem, employment, enrolled for other institute |
| Factor 4 | Institutional problem | Campus environment, too many rules, hectic schedule, low placement rate |

### Data analysis and interpretation

The preprocessed dataset was loaded into Weka data mining tool. After analyzing the dataset, the highest dropout was recorded mainly due to family reasons (51.35%) followed by institutional factors such as campus environment (21.62%), too many rules (16.22%) and low placement rate (13.51%). On the other hand, the dropout due to personal problem was mainly explained by enrollment for better prospect in other institute (37.84%), home sickness and change of goal (32.43%) followed by adjustment problem (24.32%) respectively (Table 24). However the other factors such as health problem, marriage, peer problem and hectic schedule has not played any role in discontinuing the study programme at residential university, whereas few students had reported likely to dropout due to low placement rate, hectic schedule, course difficulty and learning problem etc. Therefore, only three factors such as Family Problem, Personal Problem and Institutional Problem were considered for analysis purpose after dropping out health factors. It has been found that among the students who are affected by one factor also affected by other factors. Venn diagram (Fig. 13) shows an overall dropout picture of the students due to different factors.

**Table 24: Factors wise Dropout Reasons of Respondents in Higher Education**

| Factors | Reasons | Yes | | No | | Total students | |
|---|---|---|---|---|---|---|---|
| | | Number | percentage | Number | percentage | Number | Percentage |
| Health problem | Illness | 0 | 0.00 | 37 | 100.00 | 37 | 100.00 |
| Family Problem | Family Problem | 19 | 51.35 | 18 | 48.65 | 37 | 100.00 |
| Personal problem | Home sickness | 12 | 32.43 | 25 | 67.57 | 37 | 100.00 |
| | Marriage | 0 | 0.00 | 37 | 100.00 | 37 | 100.00 |
| | Change of personal goal | 12 | 32.43 | 25 | 67.57 | 37 | 100.00 |
| | Adjustment problem | 9 | 24.32 | 28 | 75.68 | 37 | 100.00 |
| | High course fee | 3 | 8.11 | 34 | 91.89 | 37 | 100.00 |
| | Enrolled for other institute | 14 | 37.84 | 23 | 62.16 | 37 | 100.00 |
| | Difficult Course | 1 | 2.70 | 36 | 97.30 | 37 | 100.00 |
| | Learning Problem | 1 | 2.70 | 36 | 97.30 | 37 | 100.00 |
| | Peer/friend Problems | 0 | 0.00 | 37 | 100.00 | 37 | 100.00 |
| Institutional problem | Campus environment | 8 | 21.62 | 29 | 78.38 | 37 | 100.00 |
| | Too many Rules | 6 | 16.22 | 31 | 83.78 | 37 | 100.00 |
| | Low Placement rate | 5 | 13.51 | 32 | 86.49 | 37 | 100.00 |
| | Hectic Schedule | 0 | 0.00 | 37 | 100.00 | 37 | 100.00 |

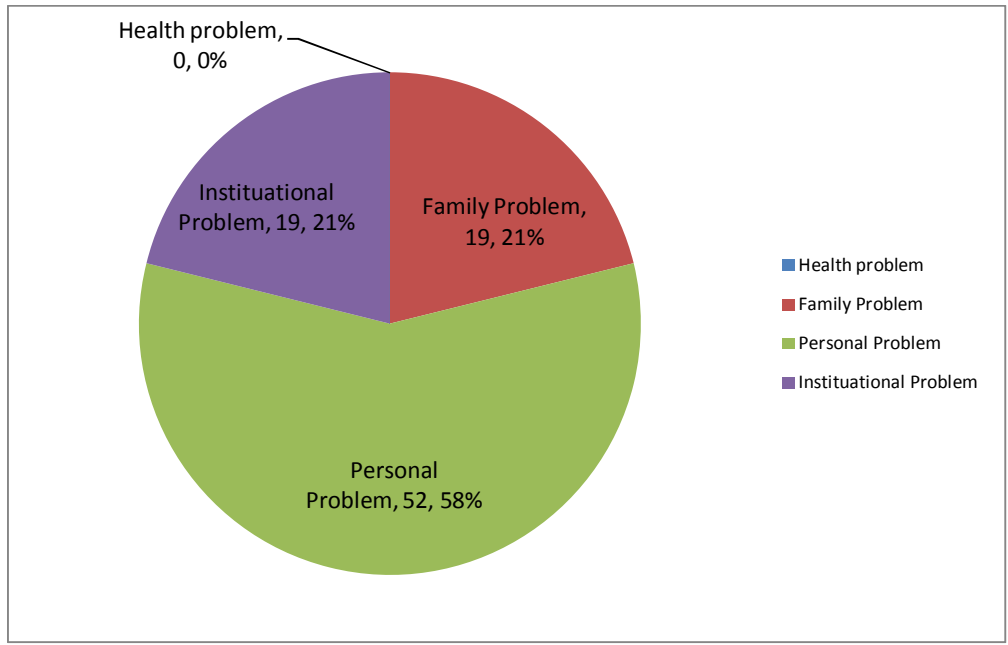

**Figure 12: Student's Dropout Factors at Residential University**

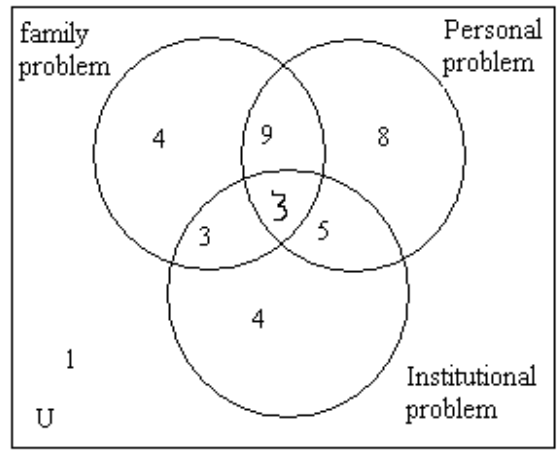

**Figure 13: Venn diagram for Dropout Factor**

Further, the tabular and binary representation of data was converted into transactional table where Family problem, Personal problem and Institutional problem denoted as F, P and I respectively (Table 25). The support for each factor was calculated on the basis of dataset. The result recorded in Table 26 shows that higher percentage of dropouts is because of the personal problem whose support value is 0.68.

**Table 25: Transaction Table**

| Transaction | Items |
|---|---|
| T1 | P |
| T2 | P |
| T3 | P,I |
| T4 | I |
| T5 | P,I |
| T6 | F,P |
| T7 | I |
| T8 | F,I |
| T9 | F |
| T10 | P,I |
| T11 | I |
| T12 | F,I |
| T13 | I |
| T14 | F,P |
| T15 | P,I |
| T16 | F,I |
| T17 | F,P |
| T18 | P |
| T19 | F |
| T20 | P |
| T21 | F |
| T22 | F,P |
| T23 | P |
| T24 | - |
| T25 | P |
| T26 | F |
| T27 | F,P |
| T28 | F,P |
| T29 | F,P |
| T30 | P |
| T31 | F,P |
| T32 | F,P,I |
| T33 | F,P,I |
| T34 | P |
| T35 | P,I |
| T36 | F,P |
| T37 | F,P,I |

**Table 26: Support Analysis**

| Dropout Factor | Support |
|---|---|
| Family | 0.51 |
| Personal | 0.68 |
| Institutional | 0.41 |
| (Family, Personal) | 0.32 |
| (Family, Institutional) | 0.16 |
| (Institutional, Personal) | 0.22 |
| (Family, Personal, Institutional) | 0.08 |

Table 27 describes that the majority of the students (63%) discontinued due to family problems were also affected by personal problems which confidence value was recorded 48%. It was found that 53% of the respondents having Institutional problem were also affected by Personal problems. On the other hand only 32 % of Personal problem students were having Institutional problems. It is interesting to note that 50% student dropout was mainly due to both family & institutional factors which also follow the personal factors. Whereas, the joint effect of family and personal factors in relation to institutional factors shows 25% student dropout. Subsequently, the joint contribution of personal and institutional factors in context with family factors explained 36% student's dropout.

**Table 27: Confidence Analysis**

| Dropout factor | Confidence |
|---|---|
| Family => Personal | Support(F,P)/support(F) =12/19 = 0.63 |
| Personal => Family | Support(P,F)/support(P) =12/25 = 0.48 |
| Institutional => Personal | Support(I,P)/support(I) = 8/15 = 0.53 |
| Personal => Institutional | Support(P,I)/support(P) = 8/25 = 0.32 |
| Family, Institutional => Personal | Support(F,I,P)/support(F,I) =3/6 =0.5 |
| Family, Personal => Institutional | Support(F,P,I)/support(F,P) =3/12 = 0.25 |
| Personal, Institutional => Family | Support(P,I,F)/support(P,I) =3/8 = 0.36 |

Overall association rule mining technique was applied to find out the relationship between two different factors affecting the student dropout at the university. From the above analysis it can be conclude that the students who have Personal problem (68%) are more prone to dropouts in comparison to Family problem (51%) and Institutional problem (41%).

# CHAPTER 5

# CONCLUSION AND FUTURE WORK

## 5.1 Conclusion

The main purpose of the study was to investigate the major factors causing the dropout of students    in undergraduate ICT Courses at Residential University.  It has been also assessed the pros cons of the affirmative action provided for the  students at the university while analyzing the reaction of respondents about university environment and university infrastructure etc.

Taking the objectives (para 1.6) into account, an extensive review of the available literature was made. Based on the review of the related literature, basic questions were formulated (see Annex-I) to indicate the nature of assumed relationships among various parameters considered in this study. To verify the stated assumptions, the study had employed different procedures and techniques. In particular, the study was conducted taking 220 samples from first year of BCA and B.Tech students of Computer department at Residential University.  Data were collected in pre-scheduled format which was handed over to the student's alongwith instructions.

Apart from its significance in providing information about the factors for students' dropout either it is low or high in higher education at residential university level, important measure could be taken to tackle it.  The present study provides valuable information to upgrade the university education system.

In order to examine the factors that affect the female students' to dropout the courses during study period and dropout reasons, both quantitative and qualitative research methods were employed. The quantitative method included the students' questionnaire concerning problems that female students face in higher education, factors that affect female students dropout and causes of attrition, teachers' and students' attitude towards affirmative action, the nature and effectiveness of affirmative action given in higher education at the residential university.

The findings of student questionnaire were analyzed and interpreted. The computer software called SPSS 20.0 was used for the treatment of the collected data. Statistical techniques such as Frequency distribution for single variable, Cross Tabulation, Chi-square test, Discriminant analysis and Data mining techniques for application of decision tree (ID3) and association rule have been used to study the causing factors for dropping out the female students.

- ➢ Based on the Correlation based feature selection result, it was found that a student's dropout was positively correlated with 12 response variables namely residence (rural/urban), family stress, family type, stream in higher secondary, satisfaction level of course, enrolled for other institute, change in goal, infrastructure of university, participation in extra-curricular activity, adjustment problem in hostel, and family problem.

- ➢ The dropout risk was observed in those students who belong to urban areas, having some stress, who belong to joint family background, opted bio or arts as a stream in higher secondary, who are not satisfied with course, applied for other institute also, whose goal was to change the stream / institution, who doesn't like the infrastructure of university and having some negative attitude towards university, who does not participated in any curricular activity, who have adjustment problem in hostel and having some problem in family.

- ➢ The association rule clearly concludes that the students who have Personal problem were more prone to dropouts in comparison to Family problem and Institutional problem. On the other hand confidence analysis shows that the majority of respondents having family problem suffered due to financial constraints and parent illness. Besides that some of the personal problem such as hostel or campus environment, unsatisfied with teaching environment and poor placement rate were forced the students to discontinue their study.

- ➢ Result indicates that ID3 decision tree algorithm is best classifier with 98% accuracy.

The generated information will be quite useful for management of university to develop policies and strategies for better planning and implementation of educational program and infrastructure under measurable condition to increase the enrolment rate in University and to take effective decision to reduce student dropout.

## 5.2 Suggestion

➢ Though the students' dropout rate is very low (<20%) which is good indicator for the university, even though the university should give due emphasis to minimize the dropout rate.

➢ The well planned and need based counseling and orientation about how to cope up with university academic life, basic life skills and possible problems that may challenge female students, should be given to female students immediately after admission.

➢ Thus, the university authorities have to organize workshops, seminars and conferences about the issues equity, affirmative action and multiculturalism to strengthen the already available positive beliefs towards affirmative action. Such type of concept will certainly reduce the student's dropout rate.

➢ Retention and university success are affected by many factors in which the institutional environment takes the lions share. Institutional factors include interaction with academic and administrative staff, gender biased institutional roles, traditional academic expectations to girls, etc. Generally, these have to be addressed by interventions like designing a gender policy at national and international level in all universities and institutions.

➢ Lastly, all higher learning institutions / universities should develop a data base about their students, teachers and administrative staff based on their background characteristics. The large dataset on student's dropout gathered by university / institution will be very much helpful to study the dropout trend at micro and macro level. Such type of study will be useful to increase the enrollment rate in various courses while updating the management system and academic curriculum etc.

## 5.3 Future Work

Future work is to study on large database of dropout student at the university using other data mining techniques such as Logistic Regression, Clustering and Neural Network in order to determine similarities and relationship between multiple factors.

# LIST OF PUBLICATIONS

## Journals

- Sweta Rai, Ajit Kumar Jain, **Student's Dropout Risk in Undergraduate Courses of ICT at Residential University- A Case Study**, IJCA International Journal of Computer Applications, Volume 84 - No 14, December 2013, ISSN: 0975-8887

- Sweta Rai, Priyanka Saini and Ajit Kumar Jain, **Model for Prediction of Dropout Student Using ID3 Decision Tree Algorithm,** International Journal of Advance Research in Computer Science & Technology (IJARCST) Vol. 2 Issue 1 Ver 2 Jan-March 2014, ISSN: 2347 -8446 (Online) ; ISSN: 2347-9817 (Print)

- Priyanka Saini, Sweta Rai and Ajit Kumar Jain, **Decision Tree Algorithm Implementation Using Educational Data,** International Journal of Computer-Aided technologies (IJCAx) Vol. 1 , No. 4, April 2014

- Priyanka Saini, Sweta Rai and Ajit Kumar Jain, **Data Mining Application in Advertisement Management of Higher Educational,** International Journal of Computer- Aided technologies (IJCAx) Vol. 1 , No. 5, April 2014

- Sweta Rai, Priyanka Saini and Ajit Kumar Jain, **Factors Affecting for Dropout Student Using Discriminant Analysis and Association Rule Mining,** submitted for International Journal of Data Mining and Emerging Technologies.

## Conferences

- Sweta Rai , Ajit Kumar Jain, **Mining Educational Data to Predict Students Dropout Factors**, "National Conference on Emerging Computing Technologies in Modern Era" to be held on January 22 2014, Submitted for National journal.

This information is only used for research purpose. The purpose of this survey is to determine the **measure factor that affects the student who is enrolled for higher technical education in the University**. Your response to this survey is crucial in providing the necessary information. The survey collects no identifying information of any respondent. All of the response in the survey will remain confidential.

## QUESTIONNAIRE

**Socio-Demographic Factors**

1. **Age**: ______        **Date of Birth:**_________

2. **Category:** General   / OBC  /  ST/SC

3. **Marital Status**: Unmarried / Married

4. **Residence**:     Rural   /   Urban

5. **Native State / State of Domicile**:

6. **Mother Tongue**: Hindi / English / Other ________

7. **Religion**:  Hindu  / Muslim/  Christian / Others(specify)________

8. **Family  Type**: Joint / Nuclear

9. **Family  Annual Income**: Below 200000 / 200001– 400000 / 400001–600000 / 600000 & above

10. **Father's Education**: Illiterate  / Secondary / Higher Secondary / Graduation / PG and above

11. **Mother's Education**: Illiterate /Secondary/ Higher Secondary / Graduation  / PG and above

12. **Father's Occupation**: Agriculture /Govt.Service / Pvt. Service / Business / Retried / NA / Others

13. **Mother's Occupation**: House wife / Govt. Service / Pvt. Service / Business / NA / Others

**Educational Factors**

14. **Grade / Percentage:  10$^{th}$** ______     **12$^{th}$** ________    **Graduation** _______

15. **Location of School education**: Village  /  Town  / City / Metropolitan city

16. **Medium of School Education**:  Hindi / English

17. **Subject stream in Higher Secondary**: Arts / Science ( Math / Bio ) / Commerce

18. **Subject stream in Graduation**: ___________

19. **Course Applied for Admission**: B. Tech / BCA / B.Sc. / M.Sc. / MCA / M. Tech / PGDCA

20. **Course Admitted:** __________

21. **Method of selection for admission** : Entrance test / Based on Marks / Management seat

22. **Year of enrollment**:_______

23. **Present Average grade point / marks**:______

24. **Satisfaction with Course:** Very satisfied / Satisfied / not very satisfied / not satisfied

25. **Syllabus of Course:** Very lengthy / lengthy / Very difficult / difficult / Satisfactory / Balanced

26. **Personal factor:**

    Illness / Family Problem / Home sickness / Campus environment / Too many rules / Low placement Rate / Course were difficult / Course fee is too high / Change of personal goal / Learning problems / others (Specify) ___________

27. **Is there any break during study?** Yes / No.
    If yes then mention reasons:

    Illness / Family Problem / Financial problem / Home sickness / Campus environment / Low placement Rate / Course were difficult / Course fee is too high / Change of personal goal / Employment / Enrolled for other institute / Learning problems / Language Barrier / Peer problems / others (Specify) ___________

**Parental factors**

28. **Whether your parents are interested for higher education?** : Yes / No.

    If **YES** then reasons: Better Educational Status / Better Marriage prospects / Economic independence / others (specify) __________

29. **Why your parent selected this university:** Due to security / It is only for women / Better educational environment / Better teaching quality / Due to reputed university / Others (specify) ________

30. **How your parents meet the university expenses:** Own income / Bank loan / Both

31. **Whether parents allow to work in private sector:** Yes / No

32. **Did your parents/ guardians:-**

    a) Encourage you to do well in university?  Yes / No

    b) Monitor and regulate yours activities?  Yes / No

    c) Provide emotional support to you?  Yes / No

    d) Show interest in your results?  Yes / No

33. **Family Experiencing stress:**

    Financial difficulties / Parent illness / others (specify) __________

**Institutional Factors**

34. **How you came to know about this university**: Friends /Newspaper / Internet / Parents / Others

35. **Did you join the university by your choice**? Yes / No

36. **Do you like this university**: Yes / No. if no then explain: _________

37. **Educational system of the university**:

    ☐ V. poor    ☐ poor    ☐ good    ☐ V. good    ☐ excellent

38. **Infrastructure of University:**

    ☐ V. poor    ☐    poor ☐    good ☐    V. go☐d excellent

39. **Extra-curriculum activities  in University:**

    ☐ V. poor    ☐ poor    ☐ good    ☐ V. good    ☐ excellent

40. **Availability of Entertainment in University campus:**

    ☐ V. poor    ☐ poor    ☐ good    ☐ V. good    ☐ excellent

41. **Is there any problem after getting enrolment in the university?**: Yes / No

    If yes then explain: _____________

42. **Attendance in class**: Regular / Irregular.  If irregular then explain:__________

43. **How much time do you spare for study:** _______________

44. **Did you participate in extra- curricular activities?** Yes / No

45. **Do you feel that you can cope with pressure at university?** Yes / No / Sometime

    If **No** then explain _______

46. **Are  you satisfied  with teacher – student relationship:**  Yes  /  No

47. **Whether university has placement cell** : Yes / No / Don't know

48. **What is  placement status** :  below average / average / good  / very good / excellent

49. **Do you want to withdraw from your course?** Yes / No / Not sure

If **YES** then plz tick mark:

☐ Financial problem   ☐ Enrolled for other institute   ☐ Low placement Rate
☐ Campus environment   ☐ Course were difficult   ☐ Course fee is too high
☐ Change of personal goal   Learning   problems                     ☐ Employment
Too many rules   Adjustment problem in hostel        Marriage        Peer /
Friend Problem              Bore                                          Other
(Specify)___________

50. **Suggestion if any to improve the educational system of the university:** _______

| S.No. | Variables | Description | Possible value |
|---|---|---|---|
| 1 | ID | A unique identification number | numeric |
| 2 | Age | Age of student in year(numeric) | numeric |
| 3 | DOB | Date of birth of student | Date( day / month / year) |
| 4 | Cat | Students category | Gen / OBC / ST/SC |
| 5 | MStatus | Marital status of student | Unmarried / Married |
| 6 | Res | Residence | Rural / Urban |
| 7 | Dom | State of Domicile ( student belong to which state ) | UP, RAJASTHAN, BIHAR, MP, UTTRAKHAND, JHARKHAND, HARYANA |
| 8 | MT | Mother tongue | Hindi / English / Other |
| 9 | Religion | Religion | Hindu / Muslim / Christian / Other |
| 10 | FType | Family Type | Joint / Nuclear |
| 11 | FAIn | Family annual income | <200000 / 200001-400000 / 400001-600000 / >600000 |
| 12 | FEdu | Father's Education | Illiterate / Secondary / Higher Secondary / Graduation / PG & above |
| 13 | MEdu | Mother's Education | Illiterate / Secondary / Higher Secondary / Graduation / PG & above |
| 14 | FOcc | Father's Occupation | Agriculture / Govt. Service / Pvt. Service / Business / Retried / NA / Other |
| 15 | MOcc | Mother's Occupation | Housewife / Govt. Service / Pvt. Service / Business / Retried / NA / Other |
| 16 | SLoc | School location of student | Village / Town / City / Metropolitian |
| 17 | HSG | Students grade / percentage in High School ($10^{th}$ ) | Numeric / Categorical |
| 18 | SSG | Students grade / percentage in Senior Secondary ($12^{th}$ ) | Numeric / Categorical |
| 19 | UGG | Students grade / percentage in Under Graduate | Numeric / Categorical |
| 20 | Med | Medium of education | Hindi / English |
| 21 | SSS | Stream in Senior Secondary($12^{th}$) | Arts / Sci.( Math / Bio) / Commerce / Home Sci. |
| 22 | SGrad | Stream in Graduation | BA / B.Sc / B.Com / BCA / BA |
| 23 | CApp | Course Applied | B.Tech / BCA / B.Sc. / M.Sc / MCA/ M.Tech / PGDCA |
| 24 | CAdm | Course Admitted | B.Tech / BCA / B.Sc. / M.Sc / MCA/ M.Tech / PGDCA |
| 25 | AType | Admission Type | Entrance / Merit / Management |
| 26 | EnrolYr | Enrollment Year of student | Numeric |
| 27 | PGrade | Present grade in Course Admitted | Numeric |
| 28 | CSatisfaction | Satisfaction with course | V.Satisfied / Satisfied / Not V.Satisfied / Not Satisfied |
| 29 | CSyllabus | Syllabus of Course | V.Lengthy / Lengthy / V. Difficult / Difficult / Satisfactory / Balanced |
| 30 | PFactor | Personal factor | Illness / Family problem / Home sickness / Campus environment / Too many rules / Low placement rate / Course were difficult / Course fee is too high / Change of personal goal / learning problem / other |

| 31 | BStudy | Break during study | Illness / Family problem / Home sickness / Campus environment / Too many rules / Low placement rate / Course were difficult / Course fee is too high / Change of personal goal / learning problem / other |
|---|---|---|---|
| 32 | IntrestHE | Parents are interested for their children's higher education | Yes / No |
| 33 | SelecUni | Why parents selected this university | Security / only for women / better edu environment / teaching quality / reputed university / other |
| 34 | UExpenses | Parents meet the University Expenses | Own Income / Bank Loan / Both |
| 35 | WPvtSector | Parents allow working in pvt. sector | Yes / No |
| 36 | FStress | Family experiences stress | Financial difficulties / illness / other |
| 37 | KnowUni | How students know about this university | Friends / Newspaper / Internet / Parents / Others |
| 38 | JoinUChoice | Join the university by your choice | Yes / No |
| 39 | LikeUni | Like this university | Yes / No |
| 40 | EduU | Educational system of the university | V.poor / poor / good / V.good / excellent |
| 41 | InfraU | Infrastructure of university | V.poor / poor / good / V.good / excellent |
| 42 | ActU | Extra-Curriculum activities in university | V.poor / poor / good / V.good / excellent |
| 43 | EntertU | Entertainment in university | V.poor / poor / good / V.good / excellent |
| 44 | Attend | Attendance in class | Regular / Irregular |
| 45 | SelfStudy | Time for self study | Numeric |
| 46 | PartActy | Participate in extra curriculum activity | Yes / No |
| 47 | UPressure | Cope with pressure at university | Yes / No / Sometime |
| 48 | PlacementCell | Whether university has placement cell | Yes / No / Not Sure |
| 49 | PlacementStatus | Placement status | Below avg. / avg. / good / v. good / excellent |
| 50 | DropoutStatus | Withdraw from your course | Yes / No / Not Sure |